\documentclass[12pt]{article}

\input epsf.sty

\newcommand{\be}[1]{ \begin{equation}\label{#1} }
\newcommand{\ee}{\end{equation}}
\newcommand{\bea}[1]{\begin{eqnarray}\label{#1} }
\newcommand{\eea}{\end{eqnarray}}

\newcommand{\p}{\partial}
\newcommand{\wt}{\widetilde}
\def\ZZZ{{\hskip-3pt\hbox{ Z\kern-1.6mm Z}}}
\def\zzz{{\hskip-3pt\hbox{ z\kern-1mm z}}}

\def\ZZZ{{\hbox{ Z\kern-1.6mm Z}}}
\def\zzz{{\hbox{ z\kern-1mm z}}}

\newcommand{\OO}{{\cal O}}

\newcommand{\EE}{{\cal E}}
\newcommand{\LL}{{\cal L}}

\newcommand{\wh}{\widehat}

\newcommand{\SSS}{{\cal S}}

\newcommand{\ben}{\begin{eqnarray}\displaystyle}
\newcommand{\een}{\end{eqnarray}}
\newcommand{\refb}[1]{(\ref{#1})}
\newcommand{\sectiono}[1]{\section{#1}\setcounter{equation}{0}}

\def\one{{\hbox{ 1\kern-.8mm l}}}
\def\zero{{\hbox{ 0\kern-1.5mm 0}}}

\newcommand{\q}{\tilde q}
\newcommand{\pp}{\tilde p}

\begin{document}

{}~
{}~

\hfill\vbox{\hbox{hep-th/0606244}
\hbox{TIFR/TH/06-15}\hbox{HRI-P-06-06-002}}\break

\vskip .4cm

\medskip

\baselineskip 20pt

\begin{center}

{\huge \bf
 Rotating  Attractors}

\end{center}

\vspace*{6.0ex}

\centerline{ \large
\rm
Dumitru Astefanesei$^a$, Kevin Goldstein$^b$,  Rudra~P.~Jena$^b$,}
\centerline{\large
\rm
Ashoke Sen$^a$ and Sandip~P.~Trivedi$^b$}

\vspace*{4.0ex}

\centerline{ \it ~$^a$Harish-Chandra Research Institute,
Chhatnag Road, Jhusi,
Allahabad 211019, India}

\vspace*{1.0ex}

\centerline{\it and}

\vspace*{1.0ex}

\centerline{  \it $^b$Tata Institute of Fundamental Research, Homi Bhabha
Road, Mumbai 400005, India}

\vspace*{5.0ex}

\centerline{\bf Abstract} \bigskip

We prove that, in a general higher derivative theory of gravity
coupled to abelian gauge fields and neutral scalar fields, the
entropy and the near horizon background of a rotating extremal black
hole is obtained by extremizing an entropy function which depends
only on the parameters labeling the near horizon background and the
electric and magnetic charges and angular momentum carried by the
black hole. If the entropy function has a unique extremum then this
extremum must be independent of the asymptotic values of the moduli
scalar fields and the solution exhibits attractor behaviour. If the
entropy function has flat directions then the near horizon
background is not uniquely determined by the extremization equations
and could depend on the asymptotic data on the moduli fields, but
the value of the entropy is still independent of this asymptotic
data. We illustrate these results in the context of two derivative
theories of gravity in several examples. These include Kerr black
hole, Kerr-Newman black hole, black holes in Kaluza-Klein theory,
and black holes in toroidally compactified heterotic string theory.

\vfill \eject

\baselineskip 18pt

\tableofcontents

\sectiono{Introduction and Summary} \label{s-1}

The attractor mechanism has played an important role in recent
studies of black holes in string theory
\cite{9508072,9602111,9602136}. According to this the geometry and
other field configurations of an extremal black hole near its
horizon is to a large extent insensitive to the asymptotic data on
the scalar fields of the theory. More precisely, if the theory
contains a set of massless scalars with flat potential --- known as
the moduli fields --- then the black hole entropy and often the near
horizon field configuration is independent of the asymptotic values
of these scalar fields.

Although initial studies of the attractor mechanism were carried out
in the context of spherically symmetric supersymmetric extremal
black holes in supergravity theories in 3+1 dimensions with two
derivative action, by now it has been generalized to many other
cases. These examples include non-supersymmetric theories, actions
with higher derivative corrections, extremal black holes in higher
dimensions etc.\cite{0507096,0510024,0511117, 0511215,0512138,
0602005,0603003,0511306,0601016,0601183,0602022,
9711053,9801081,9812082,9904005,
9906094,9910179,0007195,0009234,0012232,0409148,0410076,
0411255,0411272,0501014,
0506176,0506177,0508042,0508218,0603149,0503219,0601228,
0602292,0605279,0606108,0604106}. In particular it has been shown that in an
arbitrary theory of gravity coupled to abelian gauge fields, neutral
scalar fields and $p$-form gauge fields with a gauge and general
coordinate invariant local Lagrangian density,  the entropy of a
spherically symmetric extremal black hole remains invariant under
continuous deformation of the asymptotic data for the moduli fields
\cite{0506177,0508042}, although occasional discrete jumps are not
ruled out. In a generic situation the complete near horizon
background is independent of this asymptotic data and depends only
on the charges carried by the black hole, but in special cases
(which happen to be quite generic in supersymmetric string theories)
there may be some dependence of the near horizon background on this
asymptotic data.

Most of the studies on the attractor mechanism however have been
carried out in the context of spherically symmetric black holes
--- for some  exceptions see \cite{9602065,9611094,0503219,0605139}.
The goal of this paper is to remedy this situation and generalize
the study of the attractor mechanism to rotating black hole
solutions. Our starting point is an observation made in
\cite{9905099} that the near horizon geometries of extremal  Kerr
and Kerr-Newman black holes have SO(2,1)$\times$U(1) isometry. Armed
with this observation we prove a general result that is as powerful
as its non-rotating counterpart. In the context of 3+1 dimensional
theories, our analysis shows  that {\it in an arbitrary theory of
gravity coupled to abelian gauge fields and neutral scalar fields
with a gauge and general coordinate invariant local Lagrangian
density, the entropy of a rotating extremal black hole remains
invariant, except for occasional jumps, under continuous deformation
of the asymptotic data for the moduli fields  if an extremal black
hole is defined to be the one whose near horizon field configuration
has $SO(2,1)\times U(1)$ isometry}. In a generic situation the
complete near horizon background is independent of this asymptotic
data and depends only on the charges carried by the black hole, but
in special cases there may be some dependence of the near horizon
background on this asymptotic data.

The strategy for obtaining this result, elaborated in detail in
section \ref{s0}, is to use the entropy function formalism
\cite{0506177,0508042}. As in the case of  non-rotating black holes
we find that the near horizon background of a rotating extremal
black hole is obtained by extremizing a functional of the background
fields on the horizon, and that Wald's entropy
\cite{9307038,9312023,9403028,9502009} is given by precisely the
same functional evaluated at its extremum. Thus if this functional
has a unique extremum with no flat directions then the near horizon
field configuration is determined completely in terms of the charges
and angular momentum, with no possibility of any dependence on the
asymptotic data on the moduli fields. On the other hand if the
functional has flat directions so that the extremization equations
do not determine the near horizon background completely, then there
can be some dependence of this background on the asymptotic data,
but the entropy, being equal to the value of the functional at the
extremum, is still independent of this data. Finally, if the
functional has several extrema at which it takes different values,
then for different ranges of asymptotic  values of the moduli fields
the near horizon geometry could correspond to different extrema. In
this case as we move  in the space of asymptotic data the entropy
would change discontinuously as we cross the boundary between two
different domains of attraction, although within a given domain it
stays fixed. As in the case of non-rotating black holes, these
results are valid given the existence of a black hole solution with
SO(2,1)$\times$U(1) symmetric near horizon geometry, but our
analysis by itself does not tell us whether a solution of this form
exists. For this, one needs to carry out a more detailed analysis of
the full solution along the lines of \cite{0507096}.

Although in this paper we focus our attention on four dimensional
rotating black holes with horizons of spherical topology, the
strategy outlined above is valid for extremal
black holes in any dimension
with horizon of any compact topology, provided we define an extremal
black hole to be the one whose near horizon geometry has an
SO(2,1) isometry. The analysis is also valid for extremal black holes
in asymptotically anti de-Sitter space as long as Wald's formula for
black hole entropy
continues to hold.
In particular the proof that
the
entropy of an extremal rotating black hole in any higher derivative theory
of gravity does not change, except for occasional jumps,
under continuous variation of the asymptotic
values of the moduli fields is valid in this general context.
All that changes
is that when we try to explicitly solve the differential equations
which arise out of the extremization conditions,
we need to use boundary conditions which
are appropriate to the horizon of a given topology. Equivalently if
we carry
out the analysis by expanding various functions describing the near
horizon background in a complete set of basis functions, then we must use
basis functions which are appropriate to that given topology.
We should note however that as we vary the asymptotic values of
the moduli fields, we must hold
fixed all the conserved charges appropriate
to the particular near horizon
geometry of the black hole. This point has been elaborated further in
footnote \ref{fn2}.

In section \ref{ss2} we explore this formalism in detail in the context
of an arbitrary two derivative theory of gravity coupled to scalar and
abelian vector fields. The extremization conditions now reduce to a set
of second order differential equations with parameters and boundary
conditions which depend only on the charges and the angular momentum.
Thus the only ambiguity in the solution to these differential
equations arise from undetermined integration constants. We prove
explicitly that in  a generic situation all the integration constants are
fixed once we impose the appropriate boundary conditions and
smoothness requirement on the solutions. We also show that even in a
non-generic situation where some of the integration constants are
not fixed (and hence could depend on the asymptotic data on the moduli
fields),
the value of the entropy is independent of these undetermined
integration constants.

In section \ref{s1} we specialize even further to a class of black
holes for which all the scalar fields are constant on the horizon.
This, of course, happens automatically in theories without any
scalar fields, but also happens for purely electrically charged
black holes in theories without any $F\wt F$ type coupling in the
Lagrangian density. In this case we can solve all the differential
equations explicitly and determine the near horizon background
completely, with the constant values of the scalar fields being
determined by extremizing an effective potential --- the same
potential that appears in the determination of the attractor values
in the case of non-rotating black holes \cite{0507096}. We use these
general results to compute the entropy and near horizon geometry of
extremal Kerr as well as extremal Kerr-Newman black holes, and
reproduce the known results in these cases.

In section \ref{s3} we use a different strategy for testing our
general results. Here we take some of the known extremal rotating
black hole solutions in two derivative theories of gravity coupled
to matter, and study their near horizon geometry to determine if
they exhibit attractor behaviour. We focus on two particular classes
of examples --- the Kaluza-Klein black holes studied in
\cite{9505038,9610013,9909102} and black holes in toroidally
compactified heterotic string theory studied in \cite{9603147} (see,
also, \cite{9601118} for a restricted class of such black
holes).\footnote{Both types of black holes are special cases of
general black hole solutions in toroidally compactified heterotic
string theory and, as we show, various formul\ae\ involving entropy
and near horizon metric can be regarded as special cases of general
duality invariant formul\ae\ for these quantities.} In both these
examples, we find two kinds of extremal limits. One of these
branches, corresponding to the surface W in \cite{9505038}, does not
have an ergo-sphere and can exist only for angular momentum of
magnitude less than a certain upper bound. We call this the
ergo-free branch. The other branch, corresponding to the surface S
in \cite{9505038}, does have an ergo-sphere and can exist for
angular momentum of magnitude larger than a certain lower bound. We
call this the ergo-branch. On both branches the entropy turns out to
be independent of the asymptotic values of the moduli fields, in
accordance with our general arguments. We find however that while on
the ergo-free branch the scalar and all other background fields at
the horizon are independent of the asymptotic data on the moduli
fields, this is not the case for the ergo-branch. Thus on the
ergo-free branch we have the full attractor behaviour, whereas on
the ergo-branch only the entropy is attracted to a fixed value
independent of the asymptotic data. On general grounds we expect
that once higher derivative corrections originating at tree, loop,
and non-perturbative level are taken into account these flat
directions of the entropy function will be lifted and we shall get a
unique near horizon background even on the ergo-branch.

\sectiono{General Analysis} \label{s0}

We begin by considering a general
four dimensional theory of gravity coupled
to a set of  abelian gauge fields $A_\mu^{(i)}$
and
neutral scalar fields $\{\phi_s\}$ with action
\be{eact}
\SSS = \int d^4 x\,  \sqrt{-\det g}\,
\LL\, ,
\ee
where   $\sqrt{-\det g}\,
\LL$ is the lagrangian density,
expressed as a function of the metric $g_{\mu\nu}$,
the scalar fields
$\{\Phi_s\}$,
the gauge field strengths $F^{(i)}_{\mu\nu}=\p_\mu
A_\nu^{(i)} - \p_\nu
A_\mu^{(i)}$,
and covariant derivatives of these fields. In general $\LL$
will contain terms with more than two derivatives.
We consider a rotating extremal black hole solution whose
near horizon geometry has the symmetries of
$AdS_2\times S^1$.
The most general field configuration consistent
with the $SO(2,1)\times U(1)$ symmetry of $AdS_2\times S^1$
is of the form:
\bea{eg1}
&& ds^2\equiv g_{\mu\nu}dx^\mu dx^\nu = v_1(\theta)
\left(-r^2 dt^2+{dr^2\over
r^2}\right)  + \beta^2 \, d\theta^2+
\beta^2\, v_2(\theta) (d\phi - \alpha r dt)^2 \nonumber \\
&& \Phi_s =u_s(\theta) \nonumber \\
&& {1\over 2}\, F^{(i)}_{\mu\nu}dx^\mu\wedge dx^\nu
= (e_i-\alpha b_i(\theta)) dr \wedge dt
+ \p_\theta b_i(\theta) d\theta\wedge (d\phi - \alpha r dt)\, ,
\een
where $\alpha$, $\beta$ and $e_i$ are constants, and $v_1$, $v_2$,
$u_s$ and $b_i$  are functions of $\theta$.
Here $\phi$ is a periodic coordinate with period $2\pi$ and
$\theta$ takes value in the range $0\le\theta\le \pi$.
The SO(2,1) isometry of $AdS_2$ is generated by the Killing
vectors\cite{9905099}:
\be{ekill}
L_1=\partial_t, \qquad
L_0=t\partial_t-r\partial_r, \qquad
L_{-1}=(1/2)(1/r^2+t^2)\partial_t - (tr)\partial_r + (\alpha /r)
\partial_\phi\, .
\ee
The form
of the metric given in \refb{eg1}
implies that the black hole has zero temperature.

We shall assume that the deformed horizon, labelled by the
coordinates $\theta$ and $\phi$, is a smooth deformation
of the sphere.\footnote{Although in two derivative theories the
horizon of a four dimensional black hole is known to have
spherical topology,  once
higher derivative terms are added to the action there may be
other possibilities. Our analysis can be
easily generalized to the case where  the
horizon has the topology of
a torus rather than a sphere.
All we need is to take the $\theta$
coordinate to be a periodic variable with period $2\pi$ and expand the
various functions in the basis of periodic functions of $\theta$.
However
if the near horizon geometry is invariant under both $\phi$ and $\theta$
translations, then in the expression for $L_{-1}$ given in \refb{ekill}
we could add a term of the form $-(\gamma/ r)
\p_\theta$, and
the entropy could have an additional
dependence on the charge conjugate to the variable $\gamma$.
This represents the Noether charge
associated with $\theta$ translation, but
does not correspond to a physical charge
from the point of view of the asymptotic observer since the full solution
is not invariant under $\theta$ translation.} \label{fn2}
This requires
\bea{eg1.4}
v_2(\theta) &=& \theta^2 +\OO(\theta^4) \quad
\hbox{for $\theta\simeq 0$} \nonumber \\
&=& (\pi-\theta )^2 +\OO((\pi-\theta)^4) \quad
\hbox{for $\theta\simeq \pi$}\, .
\een
For the configuration given in
\refb{eg1} the magnetic charge associated with the $i$th gauge
field is given by
\be{eg1.5}
p_i = \int d\theta d\phi F^{(i)}_{\theta\phi} = 2\pi (b_i(\pi)
- b_i(0))\, .
\ee
Since an additive constant in $b_i$ can be absorbed into the
parameters $e_i$, we can set $b_i(0)=-p_i/4\pi$. This, together
with \refb{eg1.5}, now gives
\be{eg1.6}
b_i(0)=-{p_i\over 4\pi}, \qquad b_i(\pi) = {p_i\over 4\pi}\, .
\ee
Requiring that the gauge field strength is smooth at the north and
the south poles we get
\bea{ebcca}
b_i(\theta)&=&-{p_i\over 4\pi} +\OO(\theta^2) \quad
\hbox{for $\theta\simeq 0$} \nonumber \\
&=& {p_i\over 4\pi}
 +\OO((\pi-\theta)^2) \quad
\hbox{for $\theta\simeq \pi$}\, .
\een
Finally requiring that the near horizon
scalar fields are smooth at the poles gives
\bea{ebscalar}
u_s(\theta)&=&u_s(0) +\OO(\theta^2) \quad
\hbox{for $\theta\simeq 0$} \nonumber \\
&=& u_s(\pi)
 +\OO((\pi-\theta)^2) \quad
\hbox{for $\theta\simeq \pi$}\, .
\een
Note that the smoothness of the background requires the Taylor series
expansion around $\theta=0,\pi$ to contain only even powers of $\theta$
and $(\pi-\theta)$ respectively.

A simple way to see
the $SO(2,1)\times U(1)$ symmetry of the configuration
\refb{eg1} is
as follows.
The $U(1)$ transformation acts as a translation of $\phi$ and is
clearly a symmetry of this configuration.
In order to see the SO(2,1) symmetry of this background
we regard $\phi$ as a compact
direction and interprete this as a theory in three dimensions
labelled by coordinates $\{x^m\}\equiv(r,\theta,t)$ with
metric $\hat
g_{mn}$, vectors $a_m^{(i)}$ and $a_m$ (coming
from the $\phi$-$m$ component of the metric) and  scalar fields
$\Phi_s$,
$\psi\equiv g_{\phi\phi}$ and $\chi_i\equiv A^{(i)}_\phi$.
If we denote by $f^{(i)}_{mn}$
and $f_{mn}$ the field strengths associated with the three
dimensional gauge fields $a_m^{(i)}$ and $a_m$ respectively,
then the background \refb{eg1} can be interpreted as the following
three dimensional background:
\bea{eg2}
&& \wh{ds}^2\equiv \hat g_{mn}dx^m dx^n = v_1(\theta)
\left(-r^2 dt^2+{dr^2\over
r^2}\right)  + \beta^2 \, d\theta^2 \nonumber \\
&& \Phi_s =u_s(\theta), \qquad \psi= \beta^2\, v_2(\theta), \quad
\chi_i =   b_i(\theta)\, , \nonumber \\
&& {1\over 2}\, f^{(i)}_{mn}dx^m\wedge dx^n
= e_i\,  dr \wedge dt ,
\qquad {1\over2}\,
f_{mn}dx^m\wedge dx^n = -\alpha dr\wedge dt\, .
\een
The $(r,t)$ coordinates now describe an AdS$_2$ space and
this background is manifestly $SO(2,1)$ invariant.    In this
description the Killing vectors take the standard form
\be{ekill1}
L_1=\partial_t, \qquad
L_0=t\partial_t-r\partial_r, \qquad
L_{-1}=(1/2)(1/r^2+t^2)\partial_t - (tr)\partial_r  \, .
\ee

Eq.\refb{eg2} and hence
\refb{eg1} describes the most general field
configuration
consistent with the $SO(2,1)\times U(1)$
symmetry.
Thus in order to
derive the equations of motion we can evaluate the action on this
background and then extremize the resulting expression with respect
to the parameters labelling the background \refb{eg1}. The only
exception to this are the parameters $e_i$ and $\alpha$ labelling
the field strengths. The variation of the action with respect to these
parameters do not vanish, but give the corresponding conserved
electric charges $q_i$ and the angular momentum $J$ (which can be
regarded as the electric charge associated with the three dimensional
gauge field $a_m$.)

To implement this procedure we define:
\be{eg3}
f[\alpha,\beta,\vec e, v_1(\theta), v_2(\theta), \vec u(\theta), \vec
b(\theta)]
= \int d\theta d\phi \sqrt{-\det g}\,  \LL\, .
\ee
Note that $f$ is a function of $\alpha$, $\beta$, $e_i$
and a functional
of $v_1(\theta)$, $v_2(\theta)$,
$u_s(\theta)$ and $b_i(\theta)$. The equations
of motion now correspond
to\footnote{Our definition of the angular
momentum differs from the standard one by a $-$ sign.}
\be{eg4}
{\p f\over \p\alpha} = J, \quad {\p f\over \p\beta}=0,
\quad {\p f\over \p
e_i} = q_i\, , \quad {\delta f\over \delta v_1(\theta)}=0\, ,
\quad {\delta f\over \delta v_2(\theta)}=0,
\quad {\delta f\over \delta u_s(\theta)}=0,
\quad {\delta f\over \delta b_i(\theta)}=0\, .
\ee
Equivalently, if we define:
\be{eg5}
\EE[J,\vec q,\alpha,\beta,\vec e, v_1(\theta),
v_2(\theta), \vec u(\theta), \vec
b(\theta) ]
= 2\pi \left( J\alpha + \vec q\cdot \vec e - f[\alpha,\beta,
\vec e, v_1(\theta), v_2(\theta), \vec u(\theta), \vec
b(\theta)]
\right)\, ,
\ee
then the equations of motion take the form:
\be{eg6}
{\p \EE\over \p\alpha} = 0, \quad
{\p \EE\over \p\beta}=0, \quad {\p \EE\over \p
e_i} = 0\, , \quad {\delta \EE\over \delta v_1(\theta)}=0\, ,
\quad {\delta \EE\over \delta v_2(\theta)}=0,
\quad {\delta \EE\over \delta u_s(\theta)}=0,
\quad {\delta \EE\over \delta b_i(\theta)}=0\, .
\ee

These equations are subject to the boundary conditions \refb{eg1.4},
\refb{ebcca}, \refb{ebscalar}. For formal arguments it will be useful to
express the various
functions of $\theta$
appearing here by expanding them as a linear combination
of appropriate basis states which make the constraints \refb{eg1.4},
\refb{ebcca} manifest, and then varying $\EE$ with respect to the
coefficients appearing in this expansion. The natural functions
in terms of which we can expand an arbitrary $\phi$-independent
function on a sphere are the
Legendre polynomials $P_l(\cos\theta)$. We take
\bea{eg7}
&& v_1(\theta) = \sum_{l=0}^\infty \wt v_1(l) \, P_l(\cos\theta)\, ,
\quad v_2(\theta) = \sin^2\theta + \sin^4\theta
\sum_{l=0}^\infty \wt v_2(l) \, P_l(\cos\theta)\, , \nonumber \\
&&
u_s(\theta) = \sum_{l=0}^\infty \wt u_s(l) \, P_l(\cos\theta)\, ,
\quad b_i(\theta) = -{p_i\over 4\pi} \cos\theta
+ \sin^2\theta \sum_{l=0}^\infty \wt b_i(l) \, P_l(\cos\theta)\, .
\nonumber \\
\eea
This expansion explicitly implements the constraints \refb{eg1.4},
\refb{ebcca} and \refb{ebscalar}. Substituting this into \refb{eg5} gives
$\EE$
as a function of $J$, $q_i$, $\alpha$, $\beta$, $e_i$,
$\wt v_1(l)$, $\wt v_2(l)$, $\wt u_s(l)$ and
$\wt b_i(l)$. Thus the equations
\refb{eg6} may now be reexpressed as
 \be{eg8}
{\p \EE\over \p\alpha} = 0, \quad
{\p \EE\over \p\beta}=0, \quad {\p \EE\over \p
e_i} = 0\, , \quad {\p \EE\over \p \wt v_1(l)}=0\, ,
\quad {\p \EE\over \p \wt v_2(l)}=0,
\quad {\p \EE\over \p \wt u_s(l)}=0,
\quad {\p \EE\over \p \wt b_i(l)}=0\, .
\ee

Let us now turn to the analysis of the
entropy associated with this black hole.
For this it will be most convenient to regard this configuration
as a two dimensional extremal black hole by regarding the
$\theta$ and $\phi$ directions as compact.
In this interpretation the zero mode of the metric
$\wh g_{\alpha\beta}$ given in \refb{eg2},
with $\alpha,\beta=r,t$, is interpreted
as the two dimensional metric $h_{\alpha\beta}$:
\be{enx1}
h_{\alpha\beta} = {1\over 2} \int_0^\pi \, d\theta\, \sin\theta\,
\wh g_{\alpha\beta}\, ,
\ee
whereas all the non-zero modes of $\wh g_{\alpha\beta}$ are
interpreted as massive symmetric rank two tensor fields. This gives
\be{erel0}
h_{\alpha\beta}dx^\alpha dx^\beta =
v_1(-r^2 dt^2 + dr^2/r^2)\, , \qquad v_1 = \wt v_1(0)\, .
\ee
Thus
the near horizon configuration, regarded from two dimensions,
involves $AdS_2$ metric, accompanied by
background electric fields $f^{(i)}_{\alpha\beta}$ and
$f_{\alpha\beta}$, a set of massless and massive scalar fields
originating from the fields $u_s(\theta)$, $v_2(\theta)$
and $b_i(\theta)$, and a set of massive
symmetric rank two tensor fields originating from
$v_1(\theta)$.
According to the general results derived in
\cite{9307038,9312023,9403028,9502009}, the entropy of this
black hole is given by:
\be{ex-1}
S_{BH} = -8\pi\,   \, {\delta \SSS^{(2)}
\over \delta R^{(2)}_{rtrt}}
\, \sqrt{-h_{rr} \, h_{tt}} \,  ,
\ee
where $R^{(2)}_{\alpha\beta\gamma\delta}$ is the
two dimensional Riemann tensor associated with the metric
$h_{\alpha\beta}$, and $\SSS^{(2)}$ is the general coordinate
invariant action of this two dimensional field theory.
In taking the functional derivative with respect to
$R_{\alpha\beta\gamma\delta}$ in \refb{ex-1} we need to express all
multiple covariant derivatives in terms of symmetrized covariant
derivatives and the Riemann tensor, and then regard the components
of the Riemann tensor as independent variables.

We now note that for this two dimensional configuration
that we have,
the electric
field strengths $f^{(i)}_{\alpha\beta}$ and $f_{\alpha\beta}$ are
proportional to the volume form on $AdS_2$, the scalar fields are
constants and the tensor fields are proportional to the $AdS_2$ metric.
Thus the covariant derivatives of all
gauge and generally covariant tensors which one can
construct out of these two dimensional fields vanish.
{}In this case \refb{ex-1} simplifies to:
\be{ex0-}
S_{BH} = -8\pi\,  \sqrt{-\det h}\, {\p \LL^{(2)}
\over \p R^{(2)}_{rtrt}}
\, \sqrt{-h_{rr} \, h_{tt}} \,
\ee
where $\sqrt{-\det h}\, \LL^{(2)}$
is the two dimensional Lagrangian density, related to the four
dimensional Lagrangian density via the formula:
\be{enx2}
\sqrt{-\det h}\, \LL^{(2)} = \int d\theta d\phi \sqrt{-\det g}\,
\LL\, .
\ee
Also while computing \refb{ex0-} we set to zero
all terms in $\LL^{(2)}$ which involve covariant derivatives of the
Riemann tensor and other gauge and general coordinate
covariant combinations of fields.

We can now proceed in a manner
identical to that in \cite{0506177}
to show that the right hand side of \refb{ex0-} is the
entropy function at its extremum. First of all
from  \refb{erel0} it follows that
\be{erel1}
R^{(2)}_{rtrt} =  v_1  = \sqrt{-h_{rr} h_{tt}}\,.
\ee
Using this we can express \refb{ex0-} as
\be{erel2}
S_{BH} = -8\pi \, \sqrt{-\det h}{\p \LL^{(2)}
\over \p R^{(2)}_{rtrt}} R^{(2)}_{rtrt}\, .
\ee
Let us denote by $\LL^{(2)}_\lambda$ a deformation of
$\LL^{(2)}$ in which we replace all factors of
$R^{(2)}_{\alpha\beta\gamma\delta}$  for
$\alpha,\beta,\gamma,\delta=r, t$ by
$\lambda R^{(2)}_{\alpha\beta\gamma\delta}$, and define
\be{erel3}
f^{(2)}_\lambda \equiv \sqrt{-\det h} \, \LL^{(2)}_\lambda\, ,
\ee
evaluated on the near horizon geometry. Then
\be{erel4}
\lambda {\p f^{(2)}_\lambda\over \p\lambda} \,
= \sqrt{-\det h} \, R^{(2)}_{\alpha\beta\gamma\delta}
{\p \LL^{(2)}
\over \delta R^{(2)}_{\alpha\beta\gamma\delta}}
= 4\, \sqrt{-\det h} \, R^{(2)}_{rtrt} {\p \LL^{(2)}
\over \p R^{(2)}_{rtrt}}\, .
\ee
Using this \refb{erel2} may be rewritten as
\be{erel5}
S_{BH} = -2\pi \lambda {\p f^{(2)}_\lambda\over \p\lambda}\bigg|_{
\lambda=1}\, .
\ee

Let us now consider the effect of the scaling
\be{erel6}
\lambda \to s\lambda, \quad e_i \to s e_i, \quad \alpha\to s\alpha,
\quad \wt v_1(l)\to s \wt v_1(l)\quad \hbox{for \, \,
$0\le l<\infty$}\, ,
\ee
under which $\lambda\, R^{(2)}_{\alpha\beta\gamma\delta}\to
s^2 \, \lambda\, R^{(2)}_{\alpha\beta\gamma\delta}$. Now
since $\LL^{(2)}$ does not involve any
explicit covariant derivatives,   all indices
of $h^{\alpha\beta}$ must contract with the indices in
$f^{(i)}_{\alpha\beta}$, $f_{\alpha\beta}$, $R^{(2)}_{\alpha\beta
\gamma\delta}$ or the indices of the rank two symmetric tensor
fields whose near horizon values are given by the parameters
$\wt v_1(l)$. {}From this and
the definition of the parameters $e_i$, $\wt v_1(l)$, and
$\alpha$  it follows that $\LL^{(2)}_\lambda$ remains
invariant under
this scaling, and hence $f^{(2)}_\lambda$ transforms to
$s f^{(2)}_\lambda$, with the overall factor of $s$
coming from the $\sqrt{-\det h}$ factor in the definition of
$f^{(2)}_\lambda$. Thus we have:
\be{erel7}
\lambda {\p f^{(2)}_\lambda\over \p\lambda} + e_i
{\p f^{(2)}_\lambda\over \p e_i} + \alpha {\p f^{(2)}_\lambda\over \p
\alpha} + \sum_{l=0}^\infty \wt v_1(l)
{\p f^{(2)}_\lambda\over \p \wt v_1(l)} = f^{(2)}_\lambda\, .
\ee
Now it follows from \refb{eg3}, \refb{enx2} and \refb{erel3}
that
\be{erel8}
f[\alpha,\beta,\vec e, v_1(\theta), v_2(\theta), \vec u(\theta), \vec
b(\theta)] = f^{(2)}_{\lambda=1}\, .
\ee
Thus the extremization equations \refb{eg4} implies that
\be{erel9}
{\p f^{(2)}_\lambda\over \p e_i} = q_i, \quad
{\p f^{(2)}_\lambda\over \p
\alpha} = J, \quad {\p f^{(2)}_\lambda\over \p \wt v_1(l)}=0\, ,
\qquad \hbox{at $\lambda=1$}\, .
\ee
Hence setting $\lambda=1$ in \refb{erel7} we get
\be{erel8.5}
\lambda {\p f^{(2)}_\lambda\over \p\lambda}\bigg|_{\lambda=1}
 =
- e_i q_i - J \alpha + f^{(2)}_{\lambda=1} =
- e_i q_i - J \alpha + f[\alpha,\beta,\vec e, v_1(\theta),
v_2(\theta), \vec u(\theta), \vec
b(\theta)]\, .
\ee
Eqs.\refb{erel5} and the definition \refb{eg5} of the entropy
function  now gives
\be{ex2}
S_{BH} = \EE
\ee
at its extremum.

Using the fact that the black hole entropy is equal
to the value of the entropy
function at its extremum, we can derive some useful results following
the analysis of \cite{0506177,0508042}.
If the entropy function has a unique
extremum with no flat
directions then the extremization equations \refb{eg8} determine the
near horizon field configuration completely and the entropy as well
as the near horizon field configuration
is independent of the asymptotic moduli since the entropy function
depends only on the
near  horizon quantities.
On the other hand if the entropy function has flat directions then the
extremization equations do not determine all the near horizon parameters,
and these undetermined parameters could depend on
 the asymptotic values
of the moduli fields. However even in this case the entropy, being
independent of the flat directions, will be independent of the asymptotic
values of the moduli fields.

Although expanding various $\theta$-dependent functions in the basis
of Legendre polynomials is useful for general argument leading to
attractor behaviour, for practical computation it is often more
convenient to directly solve the differential equation in $\theta$.
For this we shall need to carefully take into account the effect of the
boundary terms. We shall see this while studying explicit examples.

 \sectiono{Extremal Rotating Black Hole in General
Two Derivative Theory} \label{ss2}

We now consider a four dimensional theory of gravity
coupled to a set of scalar fields  $\{\Phi_s\}$
and gauge fields $A_\mu^{(i)}$ with a general
two derivative action of the form:\footnote{In
the rest of the paper we shall be using the normalization of the
Einstein-Hilbert term as given in eq.\refb{et1}. This corresponds to
choosing the Newton's constant $G_N$ to be $1/16\pi$.}
\be{scation}
\SSS=\int d^4 x\, \sqrt{-\det g} \, \LL\, ,
\ee
\be{et1}
 \LL = R - h_{rs}(\vec\Phi) g^{\mu\nu}\p_\mu \Phi_s \p_\nu
\Phi_r - f_{ij}(\vec\Phi) g^{\mu\rho} g^{\nu\sigma}
F^{(i)}_{\mu\nu} F^{(j)}_{\rho\sigma} -{1\over 2}
\wt f_{ij}(\vec\Phi)\, (\sqrt{-\det g})^{-1}
\epsilon^{\mu\nu\rho\sigma} F^{(i)}_{\mu\nu} F^{(j)}_{\rho\sigma}
\, ,
\ee
where $\epsilon^{\mu\nu\rho\sigma}$ is the totally anti-symmetric
symbol with $\epsilon^{tr\theta\phi}=1$ and $h_{rs}$, $f_{ij}$ and
$\wt f_{ij}$ are fixed functions of the scalar fields $\{\Phi_s\}$.
We use the following
ansatz for the near horizon configuration of the scalar and gauge
fields\footnote{This is related to the ansatz \refb{eg1}
by a reparametrization
of the $\theta$ coordinate.}
\bea{et2}
&& ds^2 = \Omega(\theta)^2 e^{2\psi(\theta)} (- r^2 dt^2 + dr^2 / r^2 +
\beta^2 d\theta^2) + e^{-2\psi(\theta)} (d\phi-\alpha r dt)^2
\nonumber \\
&& \Phi_s =u_s(\theta) \nonumber \\
&& {1\over 2} F^{(i)}_{\mu\nu}dx^\mu\wedge dx^\nu
= (e_i-\alpha b_i(\theta)) dr \wedge dt
+ \p_\theta b_i(\theta) d\theta\wedge (d\phi - \alpha r dt)\, ,
\een
with $0\le \phi< 2\pi$, $0\le\theta\le \pi$.
Regularity at $\theta=0$ and $\theta=\pi$ requires that
\be{e1a}
\Omega(\theta) e^{\psi(\theta)} \to \hbox{constant as
$\theta\to 0,\pi$}\, ,
\ee
and
\be {e14}
\beta \Omega(\theta) e^{2\psi(\theta)} \sin\theta \to 1\quad \hbox{
as $\theta\to 0, \pi$}\, .
\ee
This gives
\bea{en1}
&& \Omega(\theta)\to a_0\sin\theta, \quad e^{\psi(\theta)}\to
{1\over \sqrt{\beta a_0} \sin\theta}, \quad \hbox{as $\theta\to 0$},
\nonumber \\
&& \Omega(\theta)\to a_\pi\sin\theta, \quad e^{\psi(\theta)}\to
{1\over \sqrt{\beta a_\pi} \sin\theta}, \quad
\hbox{as $\theta\to \pi$}\, ,
\eea
where $a_0$ and $a_\pi$ are arbitrary constants. In the next two
sections we shall describe  examples of rotating extremal black
holes in various two derivative theories of gravity with near horizon
geometry of the form described above. However
none of these black holes
will be supersymmetric even though many of them will be found
in supersymmetric theories.

Using \refb{et1}, \refb{et2} and \refb{e14} we  get
\bea{et3}
\EE &\equiv& 2\pi (J \alpha +   \vec q \cdot \vec e - \int d\theta d\phi
\sqrt{-\det g} \, \LL)
\nonumber \\
&=& 2\pi J \alpha + 2\pi \vec q \cdot \vec e -
4\pi^2 \int d\theta \, \Bigg[2\Omega(\theta)^{-1} \beta^{-1}
(\Omega'(\theta))^2 -
2\Omega(\theta) \beta - 2 \Omega(\theta) \beta^{-1}
(\psi'(\theta))^2 \nonumber \\
&&
+{1\over 2}
\alpha^2 \Omega(\theta)^{-1} \beta e^{-4\psi(\theta)}
-\beta^{-1}\Omega(\theta)
h_{rs}(\vec u(\theta))
u_r'(\theta) u_s'(\theta) + 4 \wt f_{ij}
(\vec u(\theta)) (e_i -\alpha b_i(\theta)) b_j'(\theta)\nonumber \\
&& + 2 f_{ij}(\vec u(\theta))
\left\{\beta\Omega(\theta)^{-1}
e^{-2\psi(\theta)} (e_i - \alpha b_i(\theta)) (e_j - \alpha b_j(\theta))
- \beta^{-1}\Omega(\theta) e^{2\psi(\theta)}b_i'(\theta) b_j'(\theta)
\right\}
\Bigg] \nonumber \\
&& + 8\pi^2\,   \left[ \Omega(\theta)^2 e^{2\psi(\theta)}
\sin\theta (\psi'(\theta) + 2\Omega'(\theta) / \Omega
(\theta) )\right]_{\theta=0}^{\theta=\pi}
\, .
\eea
The boundary terms in the last line of  \refb{et3} arise from
integration by parts in $\int \sqrt{-\det g}\LL$.
Eq.\refb{et3} has the property
that under a variation of $\Omega$ for which
$\delta\Omega/\Omega$ does not vanish at the boundary and/or a
variation of $\psi$ for which $\delta\psi$ does not vanish at the
boundary, the
boundary terms in $\delta\EE$
cancel if \refb{en1} is satisfied. This ensures that once the $\EE$ is
extremized under variations of $\psi$
and $\Omega$ for which $\delta\psi$ and $\delta\Omega$ vanish
at the boundary, it is also
extremized with respect to the
constants $a_0$ and $a_\pi$ appearing in \refb{en1} which changes
the boundary values of $\Omega$ and $\psi$. Also due to this
property we can now extremize the entropy function
with respect to $\beta$
without worrying about the constraint \refb{e14} since the additional
term that comes from the compensating variation in $\Omega$
and/or $\psi$ will vanish due to $\Omega$ and/or $\psi$ equations
of motion.

The equations of motion of various fields may now be obtained by
extremizing the  entropy function $\EE$ with respect
to the functions $\Omega(\theta)$, $\psi(\theta)$, $u_s(\theta)$,
$b_i(\theta)$ and the parameters $e_i$, $\alpha$,  $\beta$
labelling the near horizon geometry.
This gives
\bea{e5x}
&& - 4\beta^{-1} \Omega''(\theta)/\Omega(\theta)
+ 2 \beta^{-1} (\Omega'(\theta)/\Omega(\theta))^2 -
2\beta - 2\beta^{-1} (\psi'(\theta))^2 -{1\over 2} \alpha^2
\Omega(\theta) ^{-2} \beta
e^{-4\psi(\theta)}   \nonumber \\
&& -\beta^{-1}
h_{rs}(\vec u(\theta))
u_r'(\theta) u_s'(\theta)  \nonumber \\
&& + 2 f_{ij}(\vec u(\theta))
\left\{-\beta\Omega(\theta)^{-2}
e^{-2\psi(\theta)} (e_i - \alpha b_i(\theta)) (e_j - \alpha b_j(\theta))
- \beta^{-1}  e^{2\psi(\theta)}b_i'(\theta) b_j'(\theta)
\right\} \nonumber \\
&& = 0 \, ,
\eea
\bea{e6x}
&& 4\beta^{-1} \Omega(\theta)\psi''(\theta)
+ 4 \beta^{-1} \Omega'(\theta) \psi'(\theta) - 2\alpha^2
\Omega(\theta)^{-1} \beta e^{-4\psi(\theta)} \nonumber \\
&&
+ 2 f_{ij}(\vec u(\theta))
\left\{-2 \beta\Omega(\theta)^{-1}
e^{-2\psi(\theta)} (e_i - \alpha b_i(\theta)) (e_j - \alpha b_j(\theta))
\right.
\left. - 2 \beta^{-1}\Omega(\theta) e^{2\psi(\theta)}
b_i'(\theta) b_j'(\theta)
\right\} \nonumber \\
&&
= 0\, ,
\eea
\bea{esx1}
&& 2\left(\beta^{-1}\Omega(\theta)
h_{rs}(\vec u(\theta))
 u_s'(\theta) \right)' -\beta^{-1} \Omega(\theta) \p_r h_{ts}(\vec u
 (\theta)) u_t'(\theta) u_s'(\theta) \nonumber \\
&&
+ 2 \p_r f_{ij}(\vec u(\theta))
\left\{\beta\Omega(\theta)^{-1}
e^{-2\psi(\theta)} (e_i - \alpha b_i(\theta)) (e_j - \alpha b_j(\theta))
- \beta^{-1}\Omega(\theta) e^{2\psi(\theta)}b_i'(\theta) b_j'(\theta)
\right\} \nonumber \\
&& +4  \p_r \wt f_{ij}
(\vec u(\theta)) (e_i -\alpha b_i(\theta)) b_j'(\theta) \nonumber \\
&& =0\, ,
\eea
\bea{esx2}
&& -4 \alpha \beta f_{ij}(\vec u(\theta))
\Omega(\theta)^{-1}
e^{-2\psi(\theta)} (e_j - \alpha b_j(\theta))
+ 4 \beta^{-1} \left(f_{ij}(\vec u(\theta))
\Omega(\theta) e^{2\psi(\theta)}
b_j'(\theta)\right)'  \nonumber \\
&& \quad - 4\p_r \wt f_{ij}(\vec u(\theta)) u_r'(\theta)
(e_j - \alpha b_j(\theta)) = 0\, ,
\eea
\be{e6xa}
q_i =  8\pi \, \int\, d\theta\,
\left[ f_{ij}(\vec u(\theta)) \beta\Omega(\theta)^{-1}
e^{-2\psi(\theta)}(e_j - \alpha b_j(\theta))
+\wt f_{ij}(\vec u(\theta)) b_j'(\theta)\right]\, ,
\ee
\bea{ejx}
J &=& 2\pi \int_0^\pi d\theta\, \bigg[\alpha \Omega(\theta)^{-1}
\beta e^{-4\psi(\theta)} -4 \beta f_{ij}(\vec u(\theta))
\Omega(\theta)^{-1}
e^{-2\psi(\theta)} (e_i - \alpha b_i(\theta))   b_j(\theta)
\nonumber \\ &&
\qquad \qquad -4 \wt f_{ij}(\vec u(\theta)) b_i(\theta) b_j'(\theta)
\bigg]\, ,
\eea
\be{e7xpre}
\int d\theta\, I(\theta) = 0\, ,
\ee
\bea{e8x}
I(\theta) &\equiv& -2\Omega(\theta)^{-1} \beta^{-2} (\Omega'
(\theta))^2
-2\Omega(\theta) + 2\Omega(\theta)
\beta^{-2} (\psi'(\theta))^2 +{1\over 2} \alpha^2 \Omega(\theta)^{-1}
e^{-4\psi(\theta)}\nonumber \\
&&
+ \beta^{-2}\Omega(\theta)
h_{rs}(\vec u(\theta))
u_r'(\theta) u_s'(\theta)  \nonumber \\
&&  + 2 f_{ij}(\vec u(\theta))
\left\{ \Omega(\theta)^{-1}
e^{-2\psi(\theta)} (e_i - \alpha b_i(\theta)) (e_j - \alpha b_j(\theta))
+ \beta^{-2}\Omega(\theta) e^{2\psi(\theta)}b_i'(\theta) b_j'(\theta)
\right\}\, . \nonumber \\
\eea
Here $\prime$ denotes derivative with respect to $\theta$.
The required boundary conditions, following from the requirement
of the regularity of the solution at
$\theta=0$, $\pi$, and that the magnetic
charge vector be $\vec p$,  are:
\be{eg1.6x}
b_i(0)=-{p_i\over 4\pi},
\qquad b_i(\pi) = {p_i\over 4\pi}\, ,
\ee
\be{e1ax}
\Omega(\theta) e^{\psi(\theta)} \to \hbox{constant as
$\theta\to 0,\pi$}\, ,
\ee
\be {e14x}
\beta \Omega(\theta) e^{2\psi(\theta)} \sin\theta \to 1\quad \hbox{
as $\theta\to 0, \pi$}\, .
\ee
\be{esx0}
u_s(\theta) \to \hbox{constant as
$\theta\to 0,\pi$}\,  .
\ee
Using eqs.\refb{e5x}-\refb{esx2} one can show that
\be{e9}
I'(\theta) = 0\, .
\ee
Thus $I(\theta)$ is independent of $\theta$. As a consequence of
eq.\refb{e7xpre} we now have
\be{e7x}
I(\theta) = 0\, .
\ee

Combining eqs.\refb{e5x} and \refb{e7x} we get
\be{eky1}
\Omega'' + \beta^2 \Omega = 0\, .
\ee
A general solution to this equation is of the form
\be{e12}
\Omega = a \sin(\beta\theta + b)\, ,
\ee
where $a$ and $b$ are integration constants.
In order that $\Omega$ has the
behaviour  given in \refb{en1} for $\theta$ near 0 and $\pi$, and
not vanish at any other value of $\theta$, we must have
\be{eky2}
b=0, \qquad \beta = 1 \, ,
\ee
and hence
\be{exy2}
\Omega(\theta) = a\sin\theta\, .
\ee

In order to analyze the rest of the equations, it will be useful to
consider the Taylor series expansion of $u_r(\theta)$ and $b_i(\theta)$
around $\theta=0,\pi$
\bea{exx1}
u_r(\theta)&=& u_r(0) +{1\over 2} \theta^2 u_r''(0)+\cdots
\nonumber \\
u_r(\theta)&=& u_r(\pi)
+{1\over 2} (\theta-\pi)^2 u_r''(\pi)+\cdots
\nonumber \\
b_i(\theta) &=& b_i(0) +  {1\over 2} \theta^2 b_i''(0)
+\cdots \nonumber \\
b_i(\theta) &=& b_i(\pi)  + {1\over 2}
(\theta-\pi)^2 b_i''(\pi)
+\cdots \, ,
\eea
where we have made use of \refb{ebcca}, \refb{ebscalar}.
We now substitute \refb{exx1} into \refb{esx2} and study the
equation near $\theta=0$ by expanding the left hand side of the
equation in powers of $\theta$ and using the boundary conditions
\refb{en1}. Only odd powers of $\theta$
are non-zero.  The first non-trivial equation, appearing as the
coefficient of the order $\theta$ term, involves $b_i(0)$, $b_i''(0)$
and $b_i''''(0)$ and can be used to determine $b_i''''(0)$ in terms of
$b_i(0)$ and $b_i''(0)$.
Higher order terms determine
higher derivatives of $b_i$ at $\theta=0$
in terms of $b_i(0)$ and $b_i''(0)$.
As a result $b_i''(0)$  is not determined
in terms of $b_i(0)$
by solving the equations of motion near $\theta=0$ and
we can choose $b_i(0)$ and $b_i''(0)$
as the two independent integration constants of this equation.
Of these $b_i(0)$ is determined directly from \refb{eg1.6x}.
On the other hand for a given configuration of the other fields,
$b_i''(0)$ is also determined from \refb{eg1.6x} indirectly
by requiring that
$b_i(\pi)$ be $p_i/4\pi$. Thus we expect that generically the integration
constants associated with the solutions
to eqs.\refb{esx2} are fixed by the boundary conditions \refb{eg1.6x}.

Let us now analyze eqs.\refb{esx1} and \refb{e7x} together,  --
eq.\refb{e6x} holds automatically when the other equations are
satisfied. For this
it will be useful to introduce a new variable
\be{enewtau}
\tau =\ln\tan{\theta\over 2}\, ,
\ee
satisfying
\be{etaueq}
{d\tau\over d\theta} = {1\over \sin\theta}\, .
\ee
As $\theta$ varies from 0 to $\pi$, $\tau$ varies from $-\infty$ to
$\infty$. We denote by
$\cdot$ derivative with respect to $\tau$ and rewrite
eqs.\refb{esx1} and \refb{e7x} in this variable. This gives
\bea{esx1new}
&& 2 a^2 (h_{rs}(\vec u) \dot u_s)^\cdot
- a^2 \p_t h_{rs}(\vec u) \dot u_t \dot u_s
+4 a \p_r \wt f_{ij}
(\vec u) (e_i -\alpha b_i) \dot b_j \nonumber \\
&& \qquad + 2 \p_r f_{ij}(\vec u) \left\{
e^{-2\psi } (e_i - \alpha b_i ) (e_j - \alpha b_j )
-a^2 e^{2\psi }\dot b_i  \dot b_j\right\} =0 \, ,
\eea
and
\be{e7xnew}
-2 a^2  +2 a^2 \dot\psi^2
+ {1\over 2} \alpha^2 e^{-4\psi} + a^2 h_{rs}(\vec u) \dot u_r \dot u_s
+ 2 f_{ij}(\vec u) \left\{
e^{-2\psi } (e_i - \alpha b_i ) (e_j - \alpha b_j )
+ a^2 e^{2\psi } \dot b_i  \dot b_j
\right\} = 0\, .
\ee
If we denote by $m$ the number of scalars then we have a set of $m$
second order differential equations and one first order differential
equation, giving altogether $2m+1$
constants of integration. We want to
see in a generic situation how many
of these constants are fixed by the required
boundary conditions on $\vec u$ and $\psi$. We shall do this
by requiring that the equations and the boundary conditions are
consistent. Thus for example
if $\psi$, $\{b_i\}$ and $\{ u_s\}$ satisfy their required
boundary conditions then we can express the equations
near $\theta=0$ (or $\theta=\pi$) as:
\be{esx1ag}
2 a^2 (\hat h_{rs}  \dot u_s)^\cdot\simeq 0\, ,
\ee
and
\be{e7xag}
-2 a^2 + a^2 \hat h_{rs}
\dot u_r \dot u_s +2 a^2 \dot\psi^2\simeq 0\, .
\ee
Here
$\hat h_{rs}$ are constants giving the value of
$h_{rs}(\vec u)$ at $\vec u=\vec u(0)$ (or $\vec u=\vec u(\pi)$).
Note that we have used the boundary conditions to set some of the
terms to zero but have kept the terms containing highest derivatives
of $\psi$ and $u_r$ even if they are required to vanish due to the
boundary conditions.
The general solutions to these equations near $\theta=0$ are
\be{estart1}
u_s(\theta)\simeq c_s + v_s \tau\, , \quad
\psi(\theta) \simeq c - \tau \sqrt{1 -{1\over 2} \hat h_{rs}  v_s v_s}\, .
\ee
where $c_s$, $v_s$ and $c$ are the $2m+1$
integration constants. Since $\tau\to-\infty$ as $\theta\to 0$,
in order that $u_s$ approaches
a constant value $u_s(0)$ as $\theta\to 0$, we must require all the
$v_s$ to vanish.
On the other hand requiring that $\psi$ satisfies the boundary
condition \refb{e14x} determines $c$ to be $-\ln (2\sqrt a)$. This
gives altogether
$m+1$ conditions on the $(2m+1)$ integration constants.
Carrying out the same analysis near $\theta=\pi$ gives another
$(m+1)$ conditions among the integration constants. Thus the
boundary conditions on $\vec u$ and $\psi$ not only determine
all $(2m+1)$ integration constants of \refb{esx1new},
\refb{e7xnew}, but give an additional condition
among the as yet unknown
parameters $a$, $\alpha$ and $e_i$   entering the
equations.

This constraint, together with
the remaining equations  \refb{e6xa} and \refb{ejx}, gives
altogether $n+2$ constraints on the $n+2$ variables $e_i$, $a$
and $\alpha$, where $n$ is the number of $U(1)$ gauge fields.
Since generically $(n+2)$ equations in $(n+2)$ variables
have only a discrete number of solutions we expect that generically
the solution to eqs.\refb{e5x}-\refb{esx0} has no continuous
parameters.

In special cases however some of the integration constants
may remain undetermined, reflecting a family of solutions
corresponding to the same set of charges. As discussed in section
\ref{s0},  these represent flat directions of the entropy
function and hence the entropy associated with all members of
this family will have identical values. We shall now give a more
direct argument to this effect.
Suppose as we go from one member of the family to a
neighbouring member, each scalar field changes to
\be{d1}
u_r(\theta) \rightarrow u_r(\theta) +\delta u_r(\theta),
\ee
and suppose all the other fields and parameters
change in response, keeping the electric charges $q_i$,
magnetic charges $p_i$ and angular momentum fixed:
\bea{d2}
\Omega \rightarrow \Omega + \delta \Omega, \quad
 \psi\rightarrow \psi+ \delta \psi, \quad
 b_i \rightarrow b_i + \delta b_i, \nonumber \\
 e_i \rightarrow e_i + \delta e_i, \quad
 \alpha \rightarrow \alpha+ \delta \alpha, \quad
 \beta \rightarrow \beta + \delta \beta\, .
 \eea
Let us calculate the resulting change in the entropy $\EE$.
The changes in $e_i$,  $\alpha$, $\beta$
do not contribute to any change in $\EE$,
since $\partial_{e_i}\EE=0$,  $\partial_\alpha \EE=0$ and
$\p_\beta \EE=0$.
The only possible contributions from varying $\Omega$,
$\psi$,  $b_i$, $u_r$
can come from boundary terms, since
the bulk equations are satisfied.
Varying
$\mathcal{E}$ subject to the equations of motion,
one finds the following boundary
terms at the poles:
\begin{eqnarray}
  \delta\mathcal{E}
  & = & 8\pi^{2}
  \left[
    \beta^{-1}\Omega h_{rs}u_{r}'\delta u_{s}-2\widetilde{f}_{ij}(e_{i}-\alpha b_{i})\delta b_{j}
    +2f_{ij}\left\{ \beta^{-1}\Omega e^{2\psi}b_{i}'\right\} \delta b_{j}
  \right.\nonumber \\
  &  &
  \left.
    +\beta^{-1}\left(-2\Omega^{-1}\Omega'\delta\Omega+2\Omega\psi'\delta\psi+\delta(\Omega\psi'+2\Omega')\right)
  \right]_{\theta=0}^{\theta=\pi}\,.
  \label{eq:delta:e}
\end{eqnarray}
Terms involving $\delta b_i$ at the boundary vanish since the
boundary conditions \refb{eg1.6x}, \refb{exx1} imply that for fixed
magnetic charges $\delta b_i$ and $b_i'$ must vanish at $\theta=0$
and $\theta=\pi$. Our boundary conditions imply that variations of
$\Omega$ and $\psi$ at the poles are not independent. From the
boundary condition \refb{e14} it follows that
\begin{equation}
  \delta\Omega=-2\Omega\delta\psi
  \label{eq:delta1}
\end{equation}
at $\theta=0,\,\pi$, while from \refb{en1} one can see that at the poles
\begin{equation}
  \delta\psi'=0\,.
  \label{eq:delta2}
\end{equation}
Combining the previous two equations gives
\begin{equation}
  \delta\Omega'=-2\Omega'\delta\psi
  \label{eq:delta3}
\end{equation}
at the poles. If we vary just $\Omega$ and $\psi$ one finds
\begin{eqnarray}
  \delta_{\{\Omega,\psi\}}\mathcal{E}
  & = &
  8\pi^{2}\beta^{-1}
  \left[
    -2\Omega^{-1}\Omega'\delta\Omega+2\Omega\psi'\delta\psi+\delta(\Omega\psi'+2\Omega')
  \right]_{\theta=0}^{\theta=\pi}\nonumber \\
  & = &
  8\pi^{2}\beta^{-1}
  \left[
    4\Omega'\delta\psi+2\Omega\psi'\delta\psi+\psi'\delta\Omega+2\delta\Omega'
  \right]_{\theta=0}^{\theta=\pi} \nonumber \\
  & = & 0\, .
\end{eqnarray}
Finally,  the boundary terms proportional to $\delta u_r$ go like,
\be{d3}
\delta_{\vec u} \EE \propto \Big[\Omega h_{rs}u_r'\delta u_s
\Big]^\pi_0.
\ee
Since $\Omega \rightarrow 0$ as $\theta \rightarrow 0,\pi,$
these too vanish.
Thus we learn that the entropy is independent of any
undetermined constant of integration.

Before concluding this section we would like to note that
using the equations
of motion for various fields we can express the charges $q_i$, the
angular momentum $J$ as well as the black hole entropy, \i.e. the
value of the entropy function at its extremum, as boundary terms
evaluated at $\theta=0$ and $\theta=\pi$.
For example using \refb{esx2} we can express \refb{e6xa} as
\be{sq}
q_i={8 \pi \over \alpha}\Big[f_{ij}\Omega e^{2 \psi}
b_j'-{\tilde f}_{ij}(e_j-\alpha
 b_j)\Big]^{\theta=\pi}_{\theta=0}
\ee
Similarly using \refb{e6x} and \refb{esx2} we can express
\refb{ejx} as
\be{sj}
J ={4\pi\over \alpha}  \Big[\Omega \psi'-\Omega
f_{ij}e^{2 \psi}b_i b_j'
+  {\tilde f}_{ij}b_i (e_j-\alpha b_j)\Big]^{\theta=\pi}_{\theta=0}
-{q_i e_i \over
 2\alpha}
\ee
Finally using \refb{e5x}, \refb{e6x} we can express
the entropy function $\EE$ given in \refb{et3} as
\be{se}
\EE=8 \pi^2\left[-2 \Omega'+\Omega^2
e^{2 \psi} \sin\theta\left(\psi'+ 2 {\Omega' \over
\Omega}\right)\right]^{\theta=\pi}_{\theta=0}
\ee
Using eq.\refb{exy2} and
the boundary conditions \refb{en1} this gives,
\be{vale}
\EE=16 \pi^2 a
\ee
Using eqs. \refb{et2} and \refb{exy2}
it is easy to see that $\EE= A/4G_N$ where $A$ is the area of the
event horizon. (Note that in our conventions $G_N={1/ 16 \pi}$).
This is the expected result for theories
with two derivative action.

\sectiono{Solutions with Constant Scalars}
\label{s1}

In this section we shall solve the equations derived in section
\ref{ss2} in special cases
where there are no scalars or where the scalars
$u_s(\theta)$ are constants:
\be{ew0}
\vec u(\theta)=\vec u_0\, .
\ee
In this case
we can combine \refb{e6x}, \refb{e7x}, \refb{eky2}
and \refb{exy2}
to get
\be{epsi1}
\sin^2\theta(\psi'' + (\psi')^2) +\sin\theta\cos\theta \psi'
-{\alpha^2\over 4a^2} e^{-4\psi} -1 = 0\,  .
\ee
The unique solution to this equation subject to the boundary conditions
\refb{e14x} is:
\be{epsi2}
e^{-2\psi(\theta)} = {2a\sin^2\theta \over 2 - (1-\sqrt{1-\alpha^2})
\sin^2\theta}\, .
\ee

We now define the coordinate $\xi$ through the relation:
\be{exi1}
\xi = -{2\over \alpha} \tan^{-1} \left( {\alpha\over
1 +\sqrt{1-\alpha^2}} \cos\theta\right)\, ,
\ee
so that
\be{exi2}
d\xi = {d\theta\over \Omega(\theta)e^{2\psi(\theta)}}\, .
\ee
As $\theta$ varies from 0 to $\pi$, $\xi$ varies from $-\xi_0$
to $\xi_0$, with $\xi_0$ given by
\be{exi2.5}
\xi_0 = {1\over \alpha} \sin^{-1}\alpha\, .
\ee
In terms of this new coordinate $\xi$, \refb{esx2} takes the form:
\be{exi3}
{d^2\over d\xi^2} (e_i - \alpha b_i(\theta)) +\alpha^2
(e_i - \alpha b_i(\theta))  = 0\, .
\ee
This has solution:
\be{exi4}
(e_i - \alpha b_i(\theta)) = A_i \sin\left( \alpha \xi+B_i\right)\, ,
\ee
where $A_i$ and $B_i$ are integration constants. These can be
determined
using the boundary condition \refb{eg1.6x}:
\be{exi5}
A_i\sin(-\alpha\xi_0+B_i) = e_i +\alpha{p_i\over 4\pi}\, ,
\qquad A_i\sin(\alpha\xi_0+B_i) = e_i -\alpha{p_i\over 4\pi}\, .
\ee
This gives
\bea{ew1}
B_i &=& \tan^{-1} \left( -{4\pi e_i\over \alpha p_i}\tan(\alpha\xi_0)
\right) = \tan^{-1} \left(- {4\pi e_i\over  p_i\sqrt{1-\alpha^2}}
\right) \, , \nonumber \\
A_i &=& \left(
{e_i^2\over \cos^2(\alpha\xi_0)}+ {\alpha^2 p_i^2\over 16\pi^2
\sin^2(\alpha\xi_0)}\right)^{1/2} =
\left( {e_i^2\over 1-\alpha^2} + {p_i^2\over 16\pi^2}\right)^{1/2}\, .
\eea
Using \refb{sq} we now get:
\be{ew2}
q_i = 16 \pi \sum_j \, \left( f_{ij}(\vec u_0) \,
 \sin B_j -\wt f_{ij}(\vec u_0)\cos B_j\right) \, A_j
 = 16\, \pi \, \sum_j \left\{ f_{ij}(\vec u_0) {e_j\over \sqrt{1-\alpha^2}}
 + \wt f_{ij}(\vec u_0) \, {p_j\over 4\pi}\right\}\, .
\ee
This gives $A_i$, $B_i$ and $e_i$ in terms of $a$, $\alpha$,
$\vec u_0$
and the charges $\vec q$, $\vec p$, $J$.

Substituting the known solutions for $\Omega(\theta)$,
$\psi(\theta)$ and $b_i(\theta)$ into eq.\refb{e7x} and evaluating the
left hand side of this equation at $\theta=\pi/2$ we get
\be{enn1}
a\sqrt{1-\alpha^2} = \sum_{i,j}
f_{ij}(\vec u_0) A_i A_j \cos(B_i-B_j) =\sum_{i,j} f_{ij}(\vec u_0)
\left\{ {p_i p_j\over 16\pi^2} + {e_i e_j\over 1-\alpha^2}\right\}\, .
\ee
On the other hand \refb{sj} gives
\be{enn2}
J = 8\pi a \alpha\, .
\ee
Since $A_i$, $B_i$ and $e_i$ are known in terms of $a$, $\alpha$,
$\vec u_0$
and $\vec q$, $\vec p$, $J$, we can use \refb{enn1} and \refb{enn2}
to solve for $\alpha$ and $a$ in terms of $\vec u_0$,
$\vec q$, $\vec p$ and $J$. \refb{vale} then gives the black hole
entropy in terms of $\vec u_0$,
$\vec q$, $\vec p$ and $J$.
The final results are:
\be{alphaanew}
\alpha={J\over \sqrt{J^2 + V_{eff}(\vec u_0,\vec q, \vec p)^2}},
\qquad a = {\sqrt{J^2 + V_{eff}(\vec u_0,\vec q, \vec p)^2}\over
8\pi}\, ,
\ee
and
\be{sbhnew}
S_{BH} = 2\pi\, \sqrt{J^2 + V_{eff}(\vec u_0,\vec q, \vec p)^2}\, ,
\ee
where
\be{effpot}
V_{eff}(\vec u_0,\vec q, \vec p) = {1\over 32\pi} f^{ij}(\vec u_0)
\wh q_i \wh q_j + {1\over 2\pi} f_{ij}(\vec u_0) p_i p_j
\ee
is the effective potential introduced in \cite{0507096}.
Here $f^{ij}(\vec u_0)$ is the matrix inverse of $f_{ij}(\vec u_0)$
and
\be{ewhq}
\wh q_i \equiv q_i - 4 \, \wt f_{ij}(\vec u_0) \, p_j\, .
\ee

Finally we turn to the determination of $\vec u_0$.
If there are no scalars present in the theory then of course there
are no further equations to be solved. In the presence of scalars
we need to solve the remaining set of equations \refb{esx1}.
In the special case when all the $f_{ij}$ and $\wt f_{ij}$ are
independent of $\vec u$ these equations are satisfied by any
constant $\vec u=\vec u_0$. Thus $\vec u_0$ is undetermined and
represent flat directions of the entropy function. However if
$f_{ij}$ and $\wt f_{ij}$ depend on $\vec u$  then there will be
constraints on $\vec u_0$. First of all note that since the entropy
must be extremized with respect to all possible deformations
consistent with the $SO(2,1)\times U(1)$ symmetry, it must be
extremized with respect to $\vec u_0$. This in turn requires that
$\vec u_0$ be an extremum of $V_{eff}(\vec u_0, \vec q, \vec p)$
as in \cite{0507096}. In this case however there are further
conditions coming from \refb{esx1} since the entropy function must
also be extremized with respect to variations for which the scalar
fields are not constant on the horizon. In fact in the generic
situation it is almost impossible to satisfy \refb{esx1} with
constant $\vec u(\theta)$.
We shall now
discuss a special case where it is possible to satisfy these equations,
-- this happens
for purely electrically charged black
holes when there are no $F\wt F$ coupling in the theory (\i.e.
when $\wt f_{ij}(\vec u)=0$).\footnote{Clearly there are other
examples with non-vanishing $p_i$ and/or $\wt f_{ij}$ related
to this one by electric-magnetic duality rotation.}
In this case \refb{ew1} gives
\be{ew2.2}
B_i = {\pi\over 2}\, , \qquad A_i = {e_i\over \cos(\alpha\xi_0)}
= {e_i\over \sqrt{1-\alpha^2}}\, ,
\ee
and eqs.\refb{ew2}, \refb{exi4} give, respectively,
\be{ew2.3}
A_i = {1\over 16\pi} f^{ij}(\vec u_0) q_j, \qquad  e_i
= {\sqrt{1-\alpha^2}\over 16\pi} f^{ij}(\vec u_0) q_j\, ,
\ee
\bea{ew2.4}
(e_i - \alpha b_i(\theta)) &=& A_i\,
\cos( \alpha \xi) = {1 \over 16\pi} f^{ij}(\vec u_0) q_j
\, \cos( \alpha \xi)  \nonumber \\
&=& {1 \over 16\pi} f^{ij}(\vec u_0) q_j
\, {2\sqrt{1-\alpha^2} + (1 -\sqrt{1-\alpha^2})\sin^2\theta
\over 2 - (1 -\sqrt{1-\alpha^2})\sin^2\theta}\, .
\een
If following \refb{effpot} we now define:
\be{ev1}
V_{eff}(\vec u,\vec q) = {1\over 32\pi}\, f^{ij}(\vec u) q_iq_j\, ,
\ee
then substituting the known solutions for $\Omega$ and $\psi$
into eq.\refb{esx1} and using
\refb{ew2.4} we can see that \refb{esx1} is satisfied if
the scalars are at an extremum $\vec u_0$ of $V_{eff}$, \i.e.
\be{ev2}
\partial_r V_{eff}(\vec u_0,\vec q)=0\, .
\ee
With the help of \refb{ew2.3}, eq.\refb{enn1} now takes the form:
\be{enn3}
a\sqrt{1-\alpha^2} = {1\over 256\pi^2} f^{ij}(\vec u_0,\vec q)
q_i q_j = {1\over 8\pi} \, V_{eff}(\vec u_0,\vec q)\, ,
\ee
Using \refb{enn2}, \refb{enn3} we get
\be{etan3}
\alpha = {J\over \sqrt{J^2 +\left({V_{eff}(\vec u_0,\vec q)}
\right)^2}}\, , \qquad a = {\sqrt{J^2 +\left( V_{eff}(\vec u_0,\vec q)
\right)^2}
\over 8\pi}\, ,
\ee
\bea{etan3.5}
\Omega &=& {\sqrt{J^2 +\left( V_{eff}(\vec u_0,\vec q)
\right)^2}\over 8\pi}
\sin\theta, \nonumber \\
e^{-2\psi} &=&{1\over 4\pi}\,  {\left(J^2 +\left( V_{eff}(\vec u_0,\vec q)
\right)^2\right) \sin^2
\theta\over (1 + \cos^2\theta)
\sqrt{J^2 +\left(V_{eff}(\vec u_0,\vec q)
\right)^2}
+ V_{eff}(\vec u_0,\vec q)\sin^2\theta}\, ,
\een
\be{etan3.5.5}
(e_i - b_i(\theta)) = {1 \over 16\pi} f^{ij}(\vec u_0) q_j
\, {2 V_{eff} + (\sqrt{J^2 + V_{eff}^2} -V_{eff})\sin^2\theta
\over 2 \sqrt{J^2 + V_{eff}^2} - (\sqrt{J^2 + V_{eff}^2} -V_{eff})
\sin^2\theta}
\ee
Eq.\refb{vale} now gives the black hole entropy to be
\be{etan4}
S_{BH} = 2\pi\sqrt{J^2 +(V_{eff}(\vec u_0,\vec q)
)^2}\, .
\ee

We shall now illustrate the results using explicit examples
of  extremal Kerr black hole and extremal Kerr-Newman
black hole.

\subsection{Extremal Kerr Black Hole in Einstein Gravity}
\label{s1.1}

We consider ordinary Einstein gravity in four dimensions with
action
\be{eaction}
\SSS=\int d^4 x \sqrt{-\det g}\, \LL, \qquad \LL=R\, .
\ee
In this case since there are no matter fields we have $V_{eff}
(\vec u_0,\vec q)=0$. Let us for definiteness consider the case
where $J>0$. It then
follows from the general results
derived earlier that
 \be{enearhkn}
\alpha =1, \qquad a ={J\over 8\pi}\, ,
\ee
\be{eompsi}
\Omega = {J\over 8\pi}\,
\sin\theta, \qquad e^{-2\psi} = {J\over 4\pi} {\sin^2\theta
\over 1+\cos^2\theta}\, ,
\ee
and
\be{entkn}
S_{BH} = 2\pi J\, .
\ee
Thus determines the near horizon geometry and the entropy of an
extremal Kerr black hole and agrees with the results of \cite{9905099}.

\subsection{Extremal
Kerr-Newman Black Hole in Einstein-Maxwell Theory}
\label{s2.2}

Here we consider Einstein gravity in four dimensions coupled to a
single Maxwell field:
\be{eaknction}
\SSS=\int d^4 x \sqrt{-\det g}\, \LL, \qquad \LL= R
-{1\over 4} F_{\mu\nu} F^{\mu\nu} \, .
\ee
In this case we have $f_{11}={1\over 4}$. Hence $f^{11}=4$ and
\be{evkn}
V_{eff}
(\vec u_0,\vec q)={q^2\over 8\pi} \, .
\ee
Thus we have
\be{enearhkerr}
\alpha ={J\over \sqrt{J^2 +\left({q^2 /  8\pi}
\right)^2}}\, , \qquad a = {\sqrt{J^2 +\left({q^2 /  8\pi}
\right)^2}
\over 8\pi}\, .
\ee
\be{eomtankn}
\Omega=a\sin\theta, \qquad e^{-2\psi}
= {2a\sin^2\theta
\over 1+\cos^2\theta + q^2\sin^2\theta / \left( 8\pi
\sqrt{J^2 +\left({q^2 /  8\pi}
\right)^2}\right)}\, ,
\ee
and
\be{entkerr}
S_{BH} = 2\pi \sqrt{J^2 +\left({q^2 /  8\pi}
\right)^2}\, .
\ee
The near horizon geometry given in \refb{enearhkerr},
\refb{eomtankn} agrees with the results of \cite{9905099}.

Comparing \refb{etan3}-\refb{etan4} with
\refb{enearhkerr}-\refb{entkerr} we see that the results for the
general case of constant scalar field background is obtained from
the results for extremal  Kerr-Newman black hole carrying electric
charge $q$ via the replacement of $q$ by $q_{eff}$ where \be{eqeff}
q_{eff} = \sqrt{8\pi\, V_{eff}(\vec u_0,\vec q)}\, . \ee

\sectiono{Examples of Attractor Behaviour in Full Black Hole
Solutions} \label{s3}

The set of equations \refb{e5x}-\refb{ejx} and \refb{e7x}
are difficult to solve explicitly
in the general case. However there are many
known examples of rotating extremal black hole solutions
in a variety of
two derivative theories of gravity. In this section we shall examine
the near horizon geometry of these solutions and  check that they
obey the consequences of the generalized attractor mechanism
discussed in sections \ref{s0} and \ref{ss2}.

\subsection{Rotating Kaluza-Klein Black Holes} \label{s3.1}

In this section we consider the four dimensional theory obtained by
dimensional reduction of the five dimensional pure gravity theory
on a circle. The relevant four dimensional fields include the metric
$g_{\mu\nu}$,  a scalar field $\Phi$ associated with the radius of the
fifth dimension and a U(1) gauge field $A_\mu$. The lagrangian density
is given by
\be{ell1}
\LL = R - 2g^{\mu\nu}\p_\mu\Phi\p_\nu\Phi -e^{2\sqrt 3\Phi}
g^{\mu\rho} g^{\nu\sigma}\, F_{\mu\nu} F_{\rho\sigma} \, .
\ee
Identifying $\Phi$ as $\Phi_1$ and $A_\mu$ as $A_\mu^{(1)}$ and
comparing \refb{et1} and \refb{ell1} we see that we have in this
example
\begin{equation} \label{ehfvalues}
h_{11}=2,
 \quad f_{11}=e^{2\sqrt 3\Phi} \, .
\end{equation}

Suppose we have an extremal rotating black hole solution in this
theory with near horizon geometry of the form given in \refb{et2}.
Let us define $\tau=\ln\tan(\theta/2)$ as in \refb{enewtau},
denote by $\cdot$ derivative with respect to $\tau$ and define
\begin{equation}
\chi(\theta)=e-\alpha b(\theta)\, .
\end{equation}
Using \refb{eky2} and \refb{exy2} we can now
express appropriate linear combinations of
eqs.\refb{e6x} - \refb{esx2} and  \refb{e7x}
as
\be{dpsi}
\ddot{\psi} =
\frac{\alpha^{2}}{4a^{2}}e^{-4\psi}+1-\dot{\psi}^{2}-\dot{\Phi}^{2}
\nonumber
\ee
\begin{equation}
\ddot{\Phi}+\sqrt 3 e^{2\sqrt 3\Phi}\left\{
e^{-2\psi}a^{-2}\chi^{2}-
\alpha^{-2}e^{2\psi}\dot{\chi}^{2}\right\} =0
\label{eq:scalar} \end{equation}
\begin{equation}
\alpha^{2}a^{-2}e^{2\sqrt
3\Phi-2\psi}\chi+\left(e^{2\sqrt 3\Phi+2\psi}\dot{\chi}\right)
\dot{\,}=0\, .
\label{eq:max}
\end{equation}
\be{eieqn}
-2 a^2  +2 a^2 \dot\psi^2
+ {1\over 2} \alpha^2 e^{-4\psi} + 2 a^2   \dot \Phi^2
+  2 \left\{
e^{2\sqrt
3\Phi-2\psi}\chi^2 + a^2
\alpha^{-2} e^{2\sqrt 3\Phi+2\psi}\dot{\chi}^2
 \right\} = 0\, .
\ee
Refs.\cite{9505038,9610013,9909102} explicitly constructed
rotating charged black hole solutions in this theory.
Later we shall analyze the near
horizon geometry of these black holes in extremal limit and verify
that they satisfy eqs.\refb{dpsi}-\refb{eieqn}.

Next we note that
the lagrangian density
\refb{ell1} has a scaling symmetry:
\be{ekk1}
\Phi\to \Phi+\lambda, \qquad F_{\mu\nu} \to e^{-\sqrt 3\lambda}
F_{\mu\nu}\, .
\ee
Since the magnetic and electric
charges $p$ and $q$
are proportional to $F_{\theta\phi}$ and
$\p \LL / \p F_{rt}$ respectively, we see that under the
transformation \refb{ekk1}, $q$ and $p$ transforms to
$e^{\sqrt 3\lambda} q$ and $e^{-\sqrt 3\lambda}p$ respectively.
Thus if we want to keep the electric and the magnetic charges
fixed, we need to make a compensating transformation of the
parameters labelling the electric and magnetic charges
of the solution. This shows that we can generate a one parameter
family of solutions carrying fixed electric and magnetic charges
by using the transformation:
\be{ekk2}
\Phi\to \Phi+\lambda, \quad F_{\mu\nu} \to e^{-\sqrt 3\lambda}
F_{\mu\nu}, \quad Q\to  e^{-\sqrt 3\lambda} Q, \quad
P\to e^{\sqrt 3\lambda}P\, ,
\ee
where $Q$ and $P$ are electric and magnetic charges labelling
the original solution. This transformation will change the asymptotic
value of the scalar field $\Phi$ leaving the electric and magnetic
charges fixed. Thus according to the general
arguments given in section \ref{s0}, the entropy associated with the
solution should not change under the deformation \refb{ekk2}.
On the other hand since \refb{ekk1} is a symmetry of the theory,
the entropy is also invariant under this transformation. Combining
these two results we see that the entropy must be invariant under
\be{ekk2.5}
Q\to  e^{-\sqrt 3\lambda} Q, \quad
P\to e^{\sqrt 3\lambda}P\, .
\ee
Furthermore if the entropy function has no flat direction so that
the near horizon geometry is fixed completely by extremizing the entropy
function then the near horizon geometry, including the scalar
field configuration, should be invariant under the transformation
\refb{ekk2}.

\subsubsection{The black hole solution}

We now turn to the black hole solution described in
\cite{9505038,9610013,9909102}. The metric associated with
this solution is given by
\begin{equation} \label{eks1}
ds^{2}=-\frac{\wt\Delta}{\sqrt{f_{p}f_{q}}}
(dt-wd\phi)^{2}+\frac{\sqrt{f_{p}f_{q}}}
{\Delta}dr^{2}+\sqrt{f_{p}f_{q}}d\theta^{2}
+\frac{\Delta\sqrt{f_{p}f_{q}}}{\wt\Delta}\sin^{2}\theta d\phi^{2}
\label{eq:metric:KK}
\end{equation}
 where
\begin{eqnarray} \label{eks2}
f_{p} & = & r^{2}+a_{K}^{2}\cos^{2}\theta+r(\pp -2M_{K})
+\frac{\pp}{\pp +\q}\frac{\left(\pp -2M_{K}\right)
\left(\q-2M_{K}\right)}{2} \nonumber \\
 &  & -\frac{\pp \sqrt{\left(\pp ^{2}-4M_{K}^{2}
\right)\left(\q^{2}-4M_{K}^{2}\right)}}{2(\pp +\q)}
\frac{a_{K}}{M_{K}}\cos\theta\\
f_{q} & = & r^{2}+a_{K}^{2}\cos^{2}
\theta+r(\q-2M_{K})+\frac{\q}{\pp +\q}\frac{\left(
\pp -2M_{K}\right)\left(\q-2M_{K}\right)}{2} \nonumber \\
 & &
+\frac{\q\sqrt{\left(\pp ^{2}-4M_{K}^{2}\right)\left(\q^{2}-4M_{K}^{2}
\right)}}{2(\pp +\q)}\frac{a_{K}}{M_{K}}\cos\theta\\
w & = &
\sqrt{\pp \q}\frac{(\pp \q+4M_{K}^2)r-M_{K}(\pp -2M_{K})(\q-2M_{K})}
{2(\pp +\q)\wt\Delta}\frac{a_{K}}{M_{K}}\sin^{2}\theta\\
\Delta & = & r^{2}-2M_{K}r+a_{K}^{2}\\ \wt\Delta & = &
r^{2}-2M_{K}r+a_{K}^{2}\cos^{2}\theta \, .
\end{eqnarray}
$M_{K}$, $a_{K}$, $\pp $ and $\q$ are four parameters
 labelling the solution.
The solution for the dilaton is of the form
\begin{equation} \label{eks3}
\exp(-4\Phi/\sqrt{3})=\frac{f_{p}}{f_{q}}\, .
\end{equation}
The dilaton has been asymptotically set to
$0$, but this can be changed using the transformation \refb{ekk2}.
Finally, the gauge field is given by
\begin{equation} \label{eks4}
A_{t}=-f_{q}^{-1}\left({Q\over 4\sqrt\pi}\, \left(r+
\frac{\pp -2M_{K}}{2}\right)+\frac{1}{2}
\frac{a_{K}}{M_{K}}\sqrt{\frac{\q^{3}
\left(\pp ^{2}-4M_{K}^{2}\right)}{4\left(\pp +\q\right)}}\cos\theta\right)
\end{equation}
\begin{eqnarray}
A_{\phi} & = & -{P\over 4\sqrt\pi}\, \cos\theta-f_{q}^{-1}
{P\over 4\sqrt\pi}\, a_{K}^{2}\sin^{2}\theta\cos\theta
\nonumber \\
 &  & -\frac{1}{2}f_{q}^{-1}\sin^{2}\theta
 \frac{a_{K}}{M_{K}}\sqrt{\frac{\pp \left(\q^{2}
 -4M_{K}^{2}\right)}{4\left(\pp +\q\right)^{3}}}
 \left[ (\pp +\q)(\pp r-M_{K}(\pp -2M_{K}))+
 \q(\pp ^{2}-4M_{K}^{2})\right] \nonumber \\
\end{eqnarray}
where $Q$ and $P$, labelling the electric and magnetic charges of the
black hole, are given by,
\begin{eqnarray} \label{ePQ}
Q^{2} & = & 4\pi\, \frac{\q(\q^{2}-4M_{K}^{2})}{(\pp +\q)}\\
P^{2} & = & 4\pi\, \frac{\pp (\pp ^{2}-4M_{K}^{2})}{(\pp +\q)}\, .
\end{eqnarray}
The mass
and angular momentum of the black hole can be expressed in terms of
$M_{K}$, $a_{K}$, $\pp $ and $\q$ as follows:\footnote{In defining
the mass and angular momentum
we have taken into account the fact that we
have $G_N=1/16\pi$. At present the normalization of the charges
$Q$ and $P$ have been chosen arbitrarily, but later we shall relate
them to the charges $q$ and $p$ introduced in section \ref{ss2}.}
\begin{eqnarray} \label{eks6}
M & = & 4\pi \left(\q+\pp \right)
\end{eqnarray}
\begin{equation}
J=4\pi\,
a_{K}\, (\pp \q)^{1/2}\, \frac{\pp \q+
4M_{K}^{2}}{M_{K}(\pp +\q)}\, .
\label{eq:param:J}
\end{equation}

\subsubsection{Extremal limit: The ergo-free branch}

As first discussed in \cite{9505038}, in this case the moduli
space of extremal black holes consist of two branches.
Let us first
concentrate on one of these branches corresponding to
the surface W
in \cite{9505038}.
We consider
the  limit:  $M_{K},a_{K}\rightarrow0$
with $a_K/M_K$, $\q$ and $\pp$ held finite.
In this limit $\q$, $\pp $ and $a_K/M_K$ can
be taken as the independent parameters labelling the solution.
Then (\ref{ePQ}-\ref{eq:param:J})
become
\begin{eqnarray} \label{eks7}
M & = &  4\pi\, \left(\q+\pp \right)\\
Q^{2} & = &  4\pi\,
\frac{\q^{3}}{\left(\q+\pp \right)}\\
P^{2} & = &  4\pi\, \frac{\pp ^{3}}{\left(\q+\pp \right)}
\end{eqnarray}
\begin{equation} \label{eks8}
J=  4\pi\, {a_K\over M_K}\, {(\pp \q)^{3/2}\over
\pp+\q} = {a_K\over M_K} |PQ|\, .
\end{equation}
For definiteness we shall take $P$ and $Q$ to be positive from now on.

In this limit $\Delta$, $\wt\Delta$, $f_{p}$, $f_{q}$, $w$
and $A_\mu$ become
\begin{equation} \label{eks11}
\Delta=\wt\Delta=r^{2}
\end{equation}
\begin{eqnarray} \label{eks12}
f_{p} & = & r^{2}+\pp r+\frac{\pp ^{2}\q}{2(\pp +\q)}
\left(1-\frac{a_{K}}{M_{K}}\cos\theta\right)\\
f_{q} & = & r^{2}+\q r+
\frac{\q^{2}\pp }{2(\pp +\q)}\left(1+\frac{a_{K}}{M_{K}}
\cos\theta\right)\\
w & = & \frac{(\pp \q)^{\frac{3}{2}}}{2(\pp +\q)}
\frac{a_{K}}{M_{K}}\frac{\sin^{2}\theta}{r}= {J\over 8\pi}\,
\frac{\sin^{2}\theta}{r}
\end{eqnarray}
\begin{eqnarray} \label{eks13}
A_{t} & =& - {Q\over 4\sqrt{\pi}} f_{q}^{-1}\left(
\left(r+\frac{\pp }{2}\right)+\frac{1}{2}\left(
\frac{a_{K}}{M_{K}}\right)\pp \cos\theta\right)\\
A_{\phi} & = & -{P\over 4\sqrt{\pi}}\, \Big[\cos\theta
+\frac{1}{2}f_{q}^{-1}\sin^{2}\theta
\left(\frac{a_{K}}{M_{K}}\right)\frac{\q}{(\pp +\q)}
\left((\pp +\q)r+\q\pp \right)\Big]
\end{eqnarray}

In order that the scalar field configuration is well defined everywhere
outside the horizon, we need $f_p/f_q$ to be positive in this region.
This gives
\be{eakres}
a_K \le M_K\, .
\ee
This in turn implies that the coefficient of $g_{tt}$,
being proportional
to $\wt\Delta/\sqrt{f_p f_q}$ remains positive everywhere outside
the horizon. Thus there is no ergo-sphere for this black hole.
We call this branch of solutions  the ergo-free branch.

\subsubsection{Near horizon behaviour} \label{snearh}

In our coordinate system
the horizon is at $r=0$. To find the near horizon
geometry, we consider the limit
\begin{equation} \label{eks14}
r\rightarrow s r,\qquad t\rightarrow s^{-1}t\qquad s
\rightarrow0\, .
\end{equation}

\paragraph{Metric}
The near horizon behaviour of the metric is given by:
\begin{equation}
ds^{2}=-\frac{r^{2}}{v_{1}(\theta)}
(dt-\frac{b}{r}d\phi)^{2}+v_{1}(\theta)
\left(\frac{dr^{2}}{r^{2}}+d\theta^{2}+\sin^{2}\theta
d\phi^{2}\right)\label{eq:metric:KK:nh}\end{equation}
with
\be{eren1}
v_{1}(\theta)=\lim_{r\to 0}
\sqrt{f_pf_q} = {1\over 8\pi}\,
\sqrt{P^{2}Q^{2}-J^{2}\cos^{2}\theta},\qquad
b={J\over 8\pi}\, \sin^{2}\theta\, .
\ee
By straightforward algebraic manipulation this
metric can be rewritten as
\be{edd1}
ds^{2} = \frac{a^{2}\sin^{2}\theta}{v_{1}(\theta)}
\left(d\phi-\alpha
rdt'\right)^{2}+v_{1}(\theta)\left(-r^{2}dt'^{2}
+\frac{dr^{2}}{r^{2}}+d\theta^{2}\right)
\ee
with
\be{eren2}
t'=t/a\, ,
\ee
\be{eadef}
 a={1\over 8\pi}\, \sqrt{P^{2}Q^{2}-J^{2}}\, ,
\ee
\be{enoa}
\alpha  = -J/\sqrt{P^{2}Q^{2}-J^{2}} \, .
\ee

\paragraph{Gauge fields}

Near the horizon the gauge fields behave like
\be{eks15}
{1\over 2}\, F_{\mu\nu} dx^\mu dx^\nu =
\left[\frac{2a\sqrt\pi}{Q}
\frac{1}{\left(1+\mu\cos\theta\right)} dr \wedge dt'
+ {1\over 4\sqrt{ \pi}}\, P\sin\theta\frac{\left(1-\mu^{2}\right)}
{\left(1+\mu\cos\theta\right)^{2}} d\theta
\wedge (d\phi - \alpha r dt') \right]
\, ,
\ee
where
\be{edefmu}
\mu = {J\over PQ}\, .
\ee

\begin{figure}
\leavevmode
\begin{center}
\hbox{ 
\epsfysize=1.5in
\epsfbox{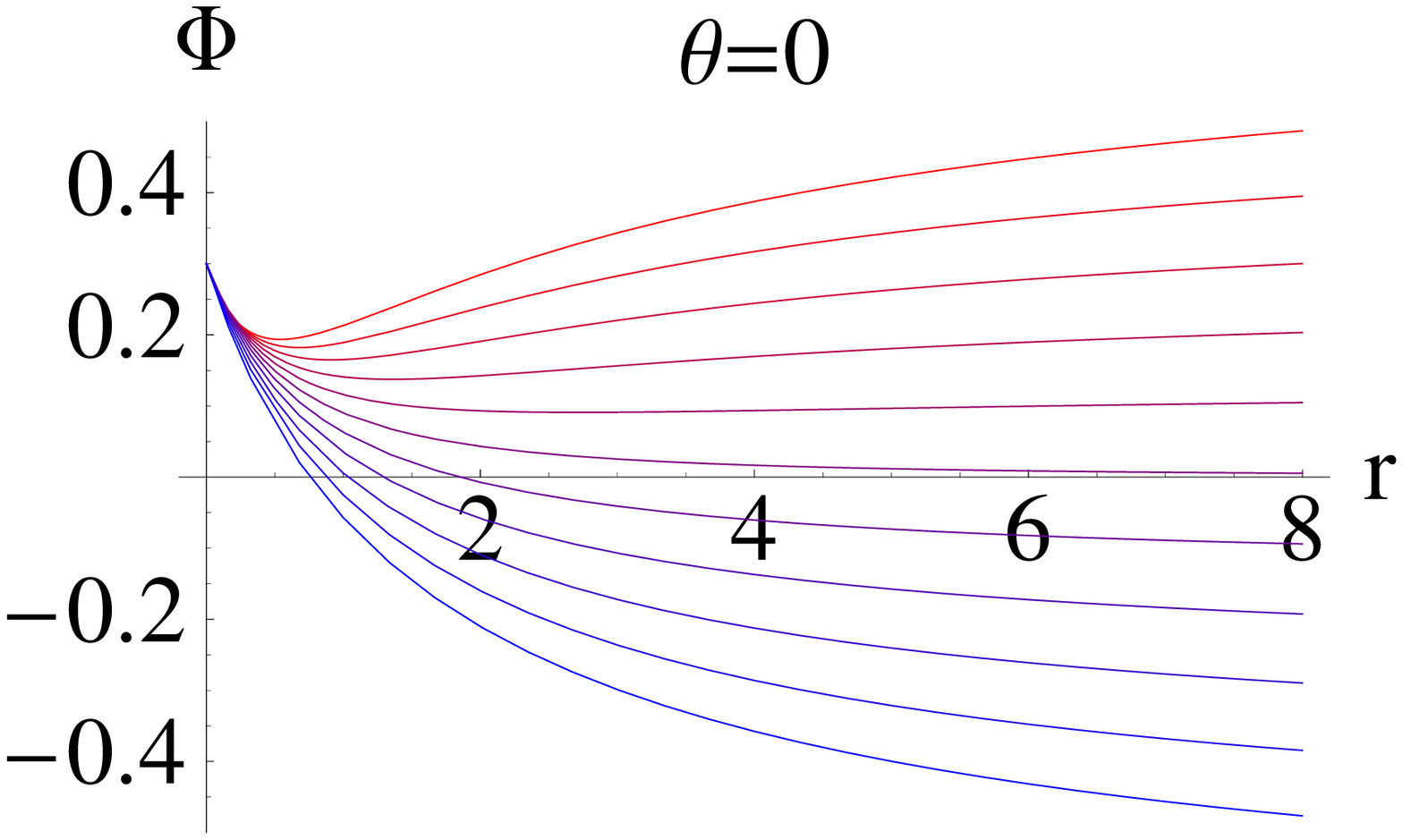}
\epsfysize=1.5in
\epsfbox{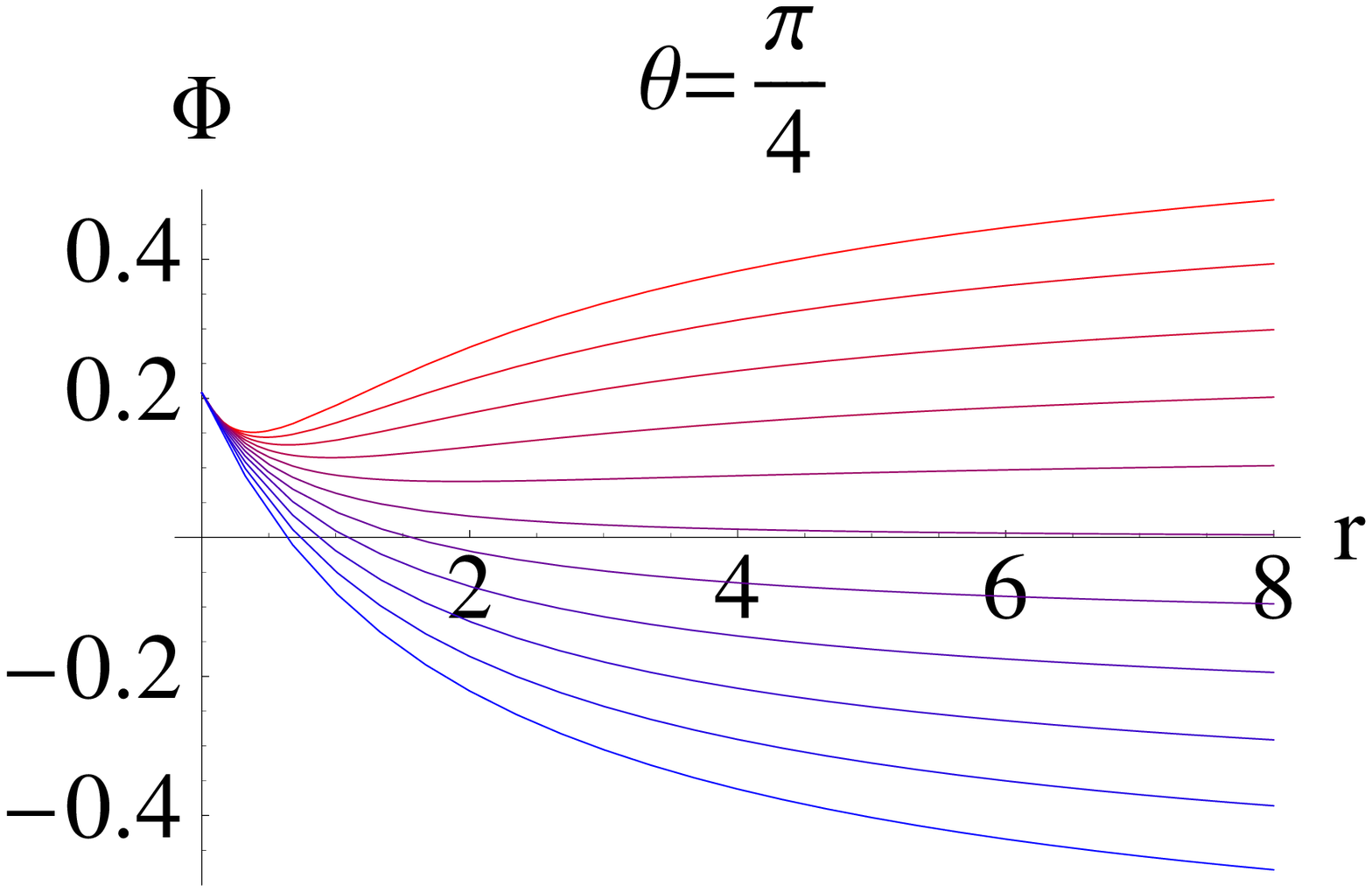}
\epsfysize=1.5in
\epsfbox{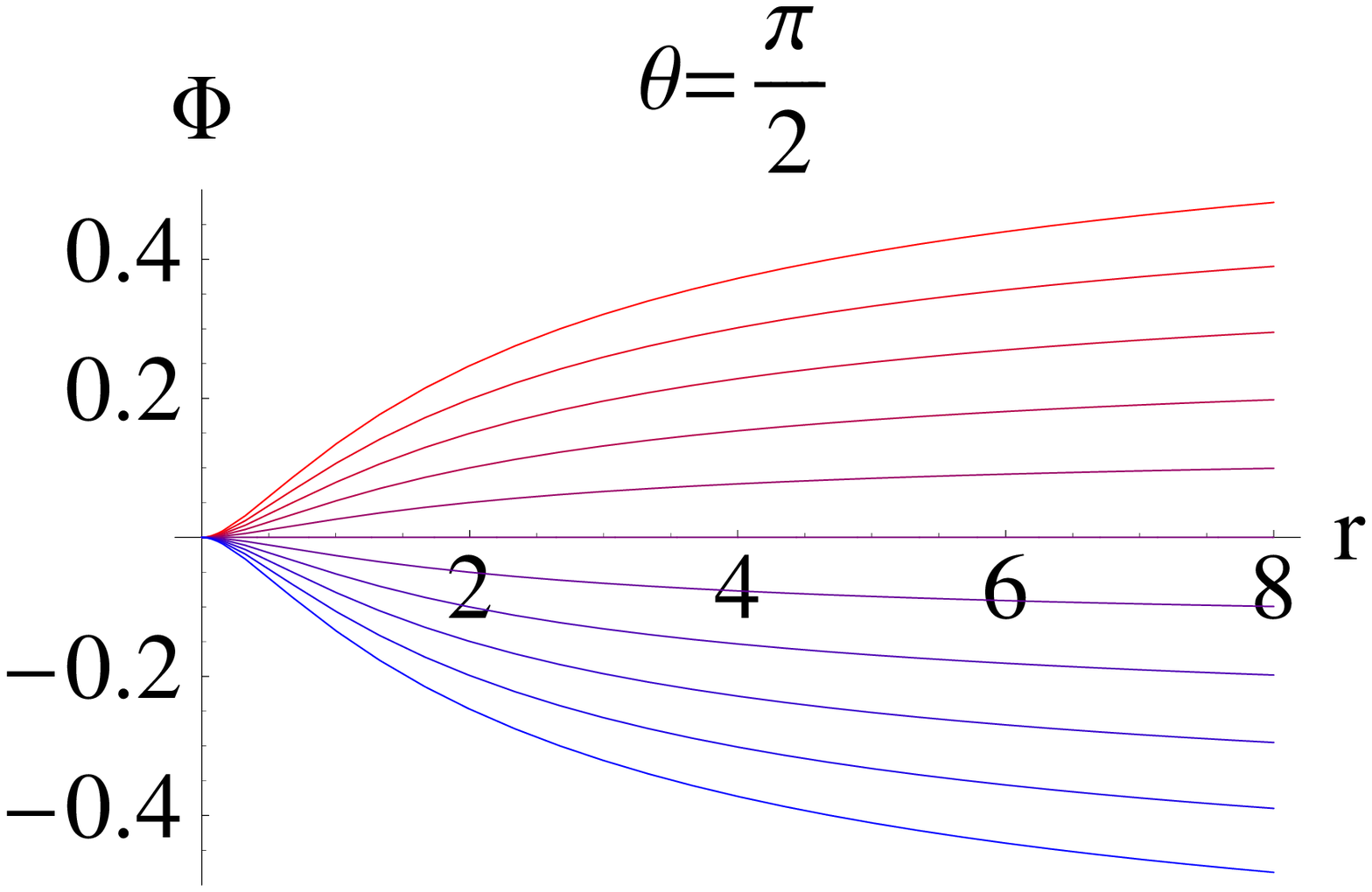}
\epsfysize=1.5in
\epsfbox{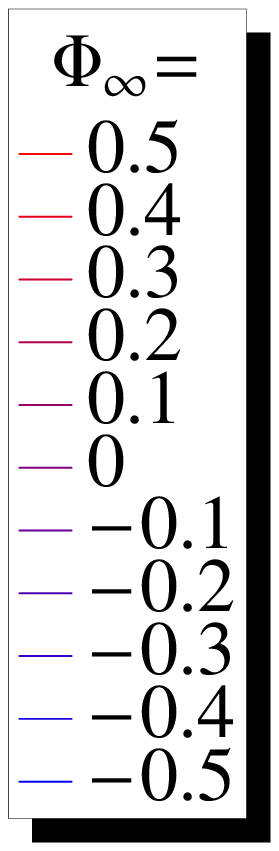}}
\end{center}
\caption{Radial evolution of the scalar field starting with different
asymptotic values at three different values of $\theta$. We take
$P=Q=4\sqrt{\pi}$, $J={16\pi/ 3}$ for $\Phi_\infty=0$, and then
change $\Phi_\infty$ and $P$, $Q$ using 
the transformation \refb{ekk2}.}
\label{f1}
\end{figure}

\paragraph{Scalar Field}

In the near horizon limit the scalar field becomes

\begin{eqnarray} \label{eks16}
\left.e^{-4\Phi/\sqrt{3}}\right|_{r=M}
& = & \left(\frac{P}{Q}\right)^{\frac{2}{3}}\frac{PQ-J\cos\theta}
{PQ+J\cos\theta}
\end{eqnarray}

\begin{figure}
\leavevmode
\begin{center}
\epsfysize=2in
\epsfbox{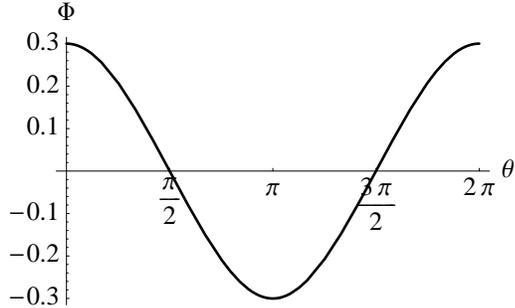}
\end{center}
\caption{Scalar field profile at the horizon of the Kaluza-Klein
black hole. We take
$P=Q=4\sqrt{\pi}$, $J={16\pi/ 3}$ for $\Phi_\infty=0$, and then
change $\Phi_\infty$ and $P$, $Q$ using 
the transformation \refb{ekk2}. The figure shows that the scalar
field profile at the horizon is independent of $\Phi_\infty$.}
\label{f2}
\end{figure}

\paragraph{Entropy}

Finally the entropy associated with this solution is
given by
\be{entr1}
S_{BH} = 4\pi \int d\theta d\phi \sqrt{g_{\theta\theta}\, g_{\phi\phi}}
= 16\pi^2 a = 2\pi \sqrt{P^2 Q^2 -J^2}\, .
\ee

We now see that the entropy is invariant under
\refb{ekk2.5} and
the near horizon background, including the scalar
field configuration given in \refb{eks16}, is invariant under the
transformation \refb{ekk2}.\footnote{As described in eqs.\refb{epexp},
\refb{qexp},
the charges $q$, $p$ are related to the parameters $Q$, $P$ by some
normalization factors. These factors do not affect the transformation laws
of the charges given in \refb{ekk2}, \refb{ekk2.5}.}
This shows that the near
horizon
field configuration is independent of the asymptotic value of the modulus
field $\Phi$. This can also be seen explicitly by studying the radial
evolution of $\Phi$ for various asymptotic values of $\Phi$; numerical
results for this evolution have been plotted in fig.\ref{f1}.
Fig.\ref{f2} shows the plot of $\Phi(\theta)$ vs. $\theta$ at the horizon
of the black hole.

\subsubsection{Entropy function analysis} \label{s3.1.3}

The analysis of section \ref{snearh} shows that
the  near horizon field configuration is precisely of the form described
in  eq.\refb{et2} with
\bea{eibi}
  \Omega(\theta) = a \sin\theta, \quad e^{-2\psi(\theta)}
= {8\pi \, a^2\sin^2\theta\over
\sqrt{P^2 Q^2 - J^2 \cos^2\theta}}\, ,
\quad
e - \alpha b(\theta) =\frac{2\sqrt \pi \, a}{Q}
\frac{1}{\left(1+\mu\cos\theta\right)}\, , \nonumber \\
e^{-4\Phi/\sqrt 3} =
\left(\frac{P}{Q}\right)^{\frac{2}{3}}\frac{PQ-J\cos\theta}
{PQ+J\cos\theta}\, , \quad
a  = {1\over 8\pi}\, \sqrt{P^{2}Q^{2}-J^{2}}, \quad
\alpha = -{J\over \sqrt{P^{2}Q^{2}-J^{2}}} \, . \nonumber \\
\eea
We can easily verify that this configuration satisfies
eqs.\refb{dpsi}-\refb{eieqn} obtained by
extremizing the entropy function.

Using eq.\refb{eg1.6x} with values of $h_{11}$ and $f_{11}$ given
in \refb{ehfvalues} we  get
\be{eeexp}
e = {1\over 2} \left[(e-\alpha b(\pi)) + (e-\alpha b(0))\right]
= {P^2 Q\over 4\sqrt\pi \sqrt{P^2 Q^2 - J^2}}\, ,
\ee
and
\be{epexp}
p = -{2\pi\over  \alpha} \left[(e-\alpha b(\pi)) - (e-\alpha b(0))\right]
= \sqrt\pi\, P\, .
\ee
Eq.\refb{sq} now gives
\be{qexp}
q = {8\pi\over \alpha} \, \left[ {e^{2\sqrt 3\Phi} b'\over
\sin\theta}\right]_0^\pi = 4\sqrt\pi\, Q\, .
\ee
Finally the right hand side of eq.\refb{sj} evaluated for the background
\refb{eibi} gives the answer $J$ showing that we have correctly identified
the parameter $J$ as the angular momentum carried by the black hole.

\subsubsection{The ergo-branch} \label{sbranch}

 The extremal limit on this branch, corresponding to the
surface S in \cite{9505038}, amounts to taking
  \be{esb1}
 a_K=M_K
 \ee
 in the black hole solution. Thus we have the relations
 \be{esb2}
 Q^{2} = 4\pi\, \frac{\q(\q^{2}-4M_{K}^{2})}{(\pp +\q)}, \quad
P^{2} = 4\pi\, \frac{\pp (\pp ^{2}-4M_{K}^{2})}{(\pp +\q)}, \quad
J= 4\pi\, \sqrt{\pp \q} \, \frac{\pp \q +4M_{K}^{2}
}{(\pp +\q)}\, .
\ee
In order to take the near horizon
limit of this solution we first let
\begin{equation}
  \label{eq:new:r}
  r\rightarrow r+M_K
\end{equation}
which shifts the horizon to $r=0$. Near the horizon
$\Delta$, $\tilde\Delta$ and $w$ become
\begin{eqnarray}
  \label{eq:deltas}
  \Delta &=& r^2 \\
  \tilde\Delta &=&  -M_K^2\sin^2\theta + \OO(r^2) \\
  w &=& -{\sqrt{\q\pp}}\, (1+\bar w r ) + \OO(r^2)
\end{eqnarray}
with
\begin{equation}
  \label{eq:wbar}
  \bar w = {\pp \q+4M_K^2\over 2(\pp+\q)M_K^2}\, .
\end{equation}
Note that $\wt\Delta$ changes from being positive at large
distance to  negative at the horizon. Thus $g_{tt}$ changes
sign as we go from the asymptotic region to the horizon and the
solution has an ergo-sphere. We call this branch of solutions  the
ergo-branch. Using eqs.\refb{eq:deltas}-\refb{eq:wbar}
we can  write the  metric as
\begin{equation} \label{nh:s:1}
  ds^{2}=\frac{M_K^2\sin^2\theta}{\sqrt{f_p f_q}}\left(dt+
{\sqrt{\q\pp}}(1+\bar w r)d\phi\right)^{2}
+{\sqrt{f_p f_q}}\left({dr^{2}\over r^2}
+d\theta^{2}-{r^2\over M_K^2}
d\phi^{2}\right) + \cdots
\end{equation}
where $\cdots$ denote terms which
will eventually vanish in the near horizon limit that we are
going to describe below.
After letting
 \begin{equation}
   \label{eq:phi:shift}
   \phi\rightarrow\phi-t/\sqrt{\q\pp}
 \end{equation}
 and taking the near horizon limit
 \be{sblimit}
 r\to s\, r, \qquad t\to s^{-1} \, t, \qquad s\to 0\, ,
 \ee
the metric  becomes
\begin{equation} \label{nh:s:2}
  ds^{2}=\frac{M_K^2\sin^2\theta}{v_1(\theta)}(\sqrt{\q\pp}
  d\phi-\bar w rd t)^{2}
+{v_1(\theta)}\left({dr^{2}\over r^2}
+d\theta^{2}-{r^2\over  M_K^2\, {\q\pp}} dt^{2}\right)
\end{equation}
where
\be{edefv1}
v_1(\theta) = \lim_{r\to 0} \sqrt{f_p\, f_q}\, .
\ee
Finally rescaling
\begin{equation}
  \label{eq:t:shift}
  t\rightarrow M_K \sqrt{\q\pp}~ t
\end{equation}
the metric becomes of the form given in \refb{et2} with
\be{esb3}
\Omega=M_K\sqrt{\pp \q}\, \sin\theta, \quad e^{-2\psi} =
{M_K^2 \pp \q\sin^2\theta \over v_1(\theta)}, \quad \alpha=
M_K ~\bar w\, .
\ee
Using  eqs.\refb{eq:wbar} and \refb{esb2}
we find that
\begin{equation}
  \label{eq:w:vs:alpha}
  \alpha = {J\over\sqrt{J^2-P^2 Q^2}}, \quad \Omega
  ={1\over 8\pi} \, \sqrt{J^2 - P^2 Q^2} \, \sin\theta, \quad
  e^{-2\psi} = {(J^2 - P^2 Q^2) \sin^2\theta \over 64 \pi^2
  v_1(\theta)}\, .
\end{equation}

\begin{figure}
\leavevmode
\begin{center}
\hbox{\epsfysize=1.25in
\epsfbox{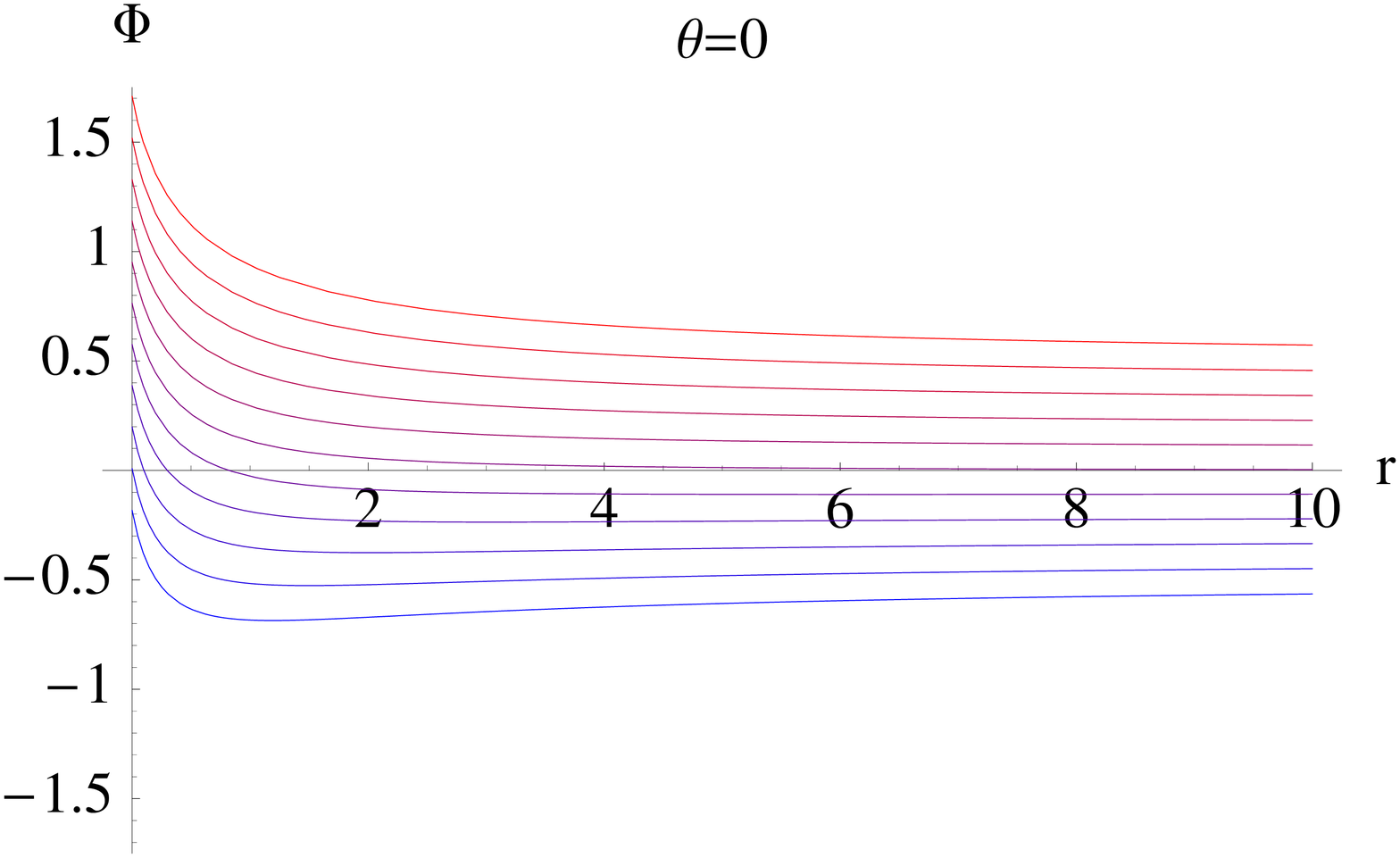}
\epsfysize=1.25in
\epsfbox{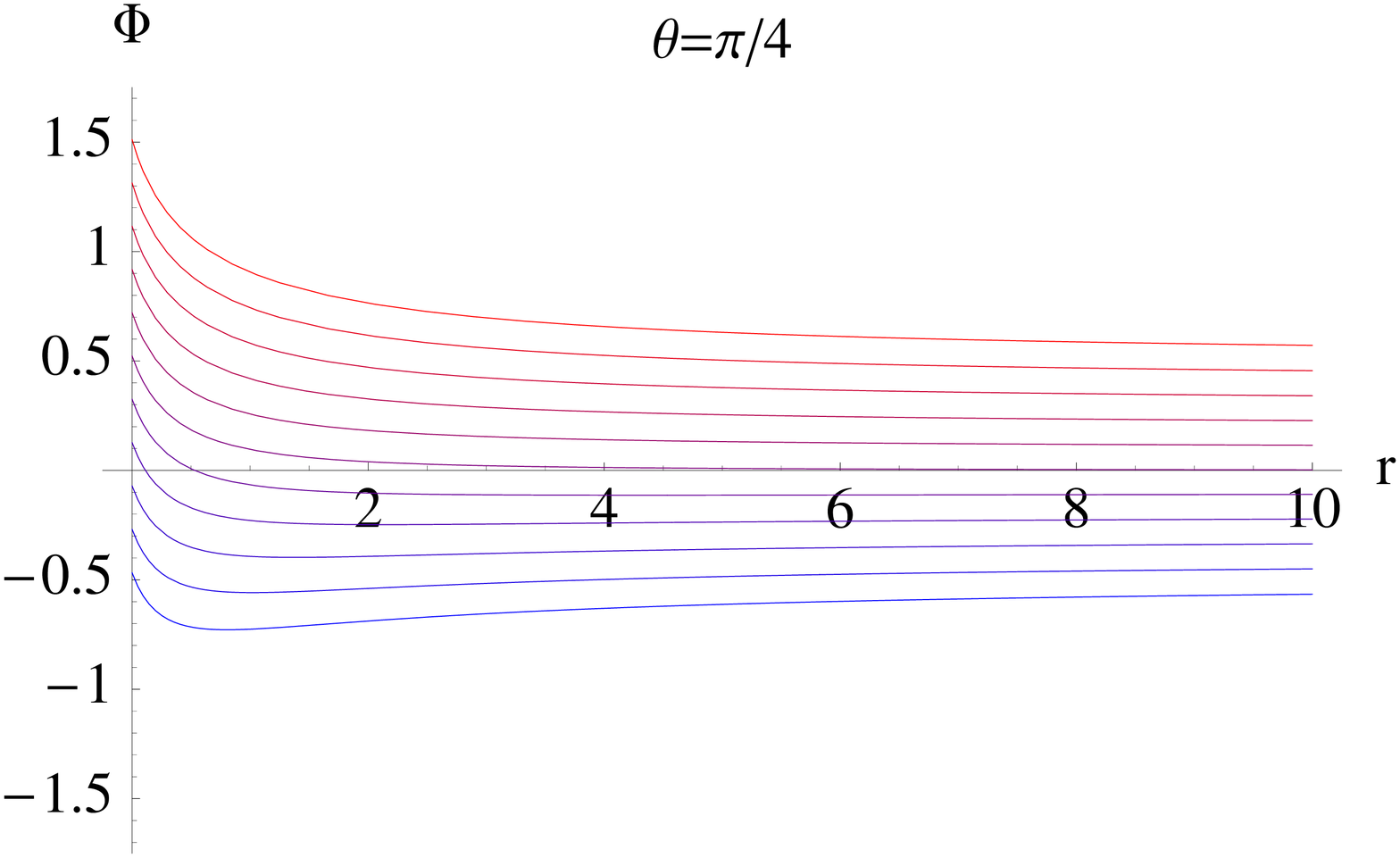}
\epsfysize=1.25in
\epsfbox{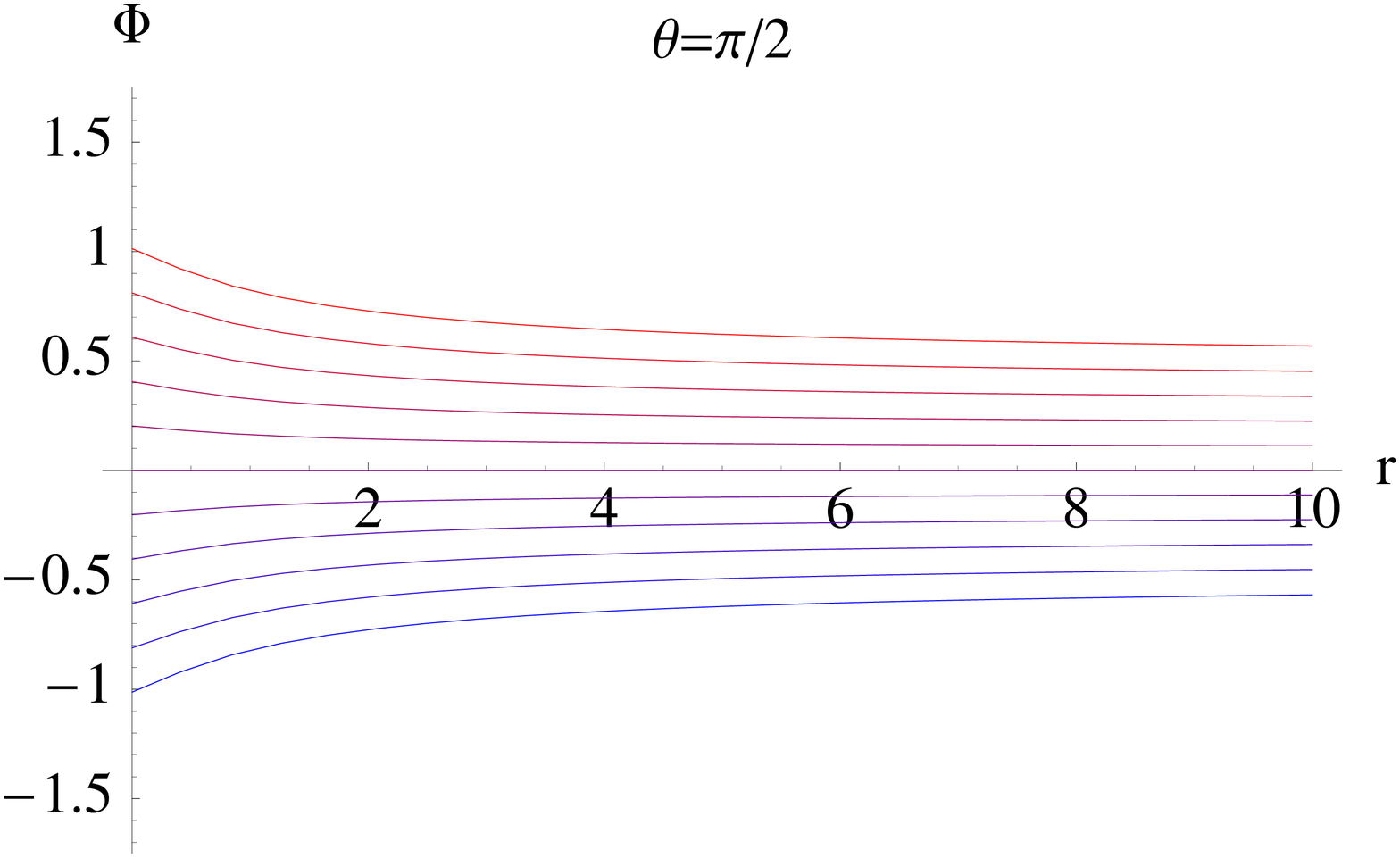}}
\hbox{\epsfysize=1.25in
\epsfbox{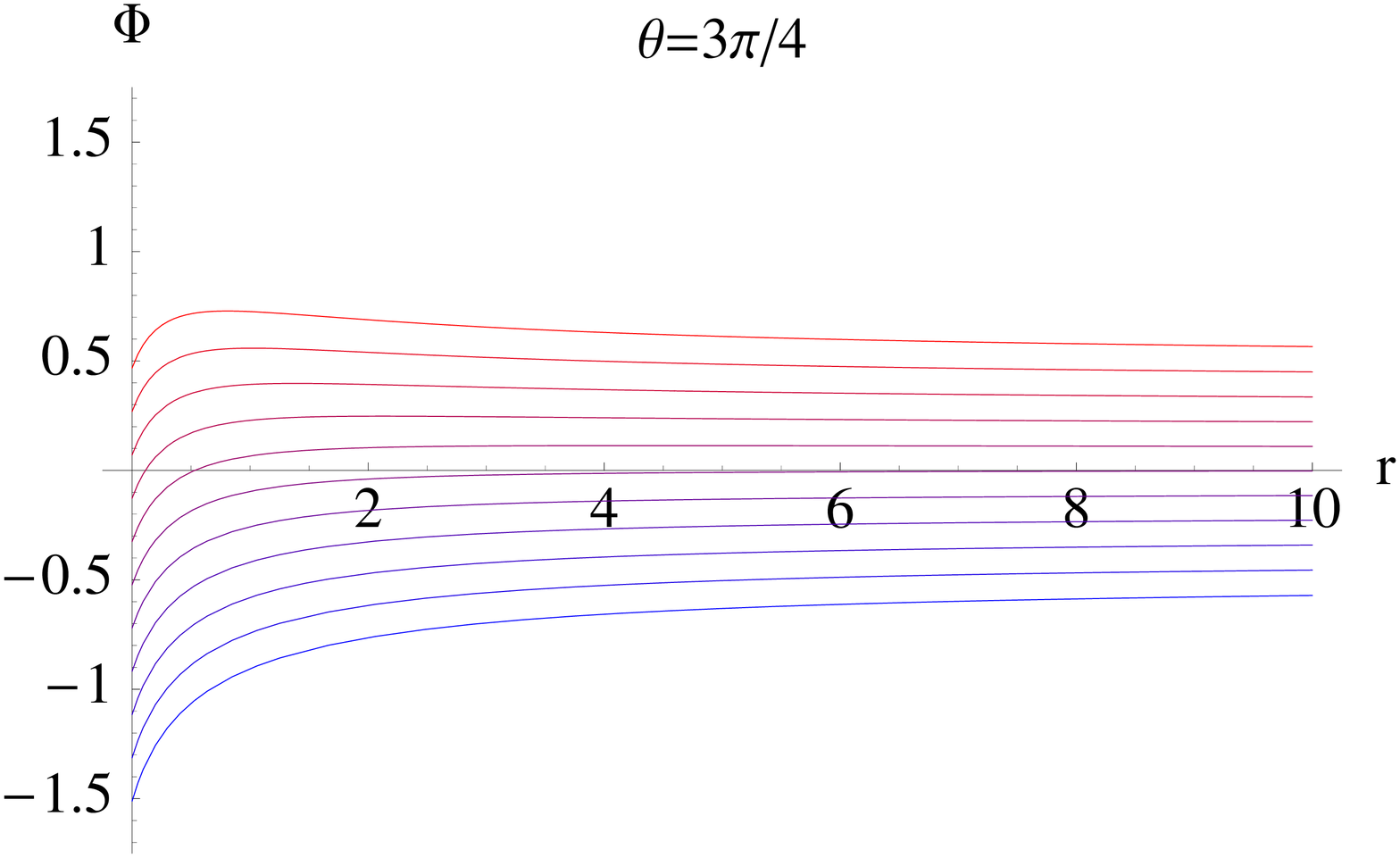} 
\epsfysize=1.25in
\epsfbox{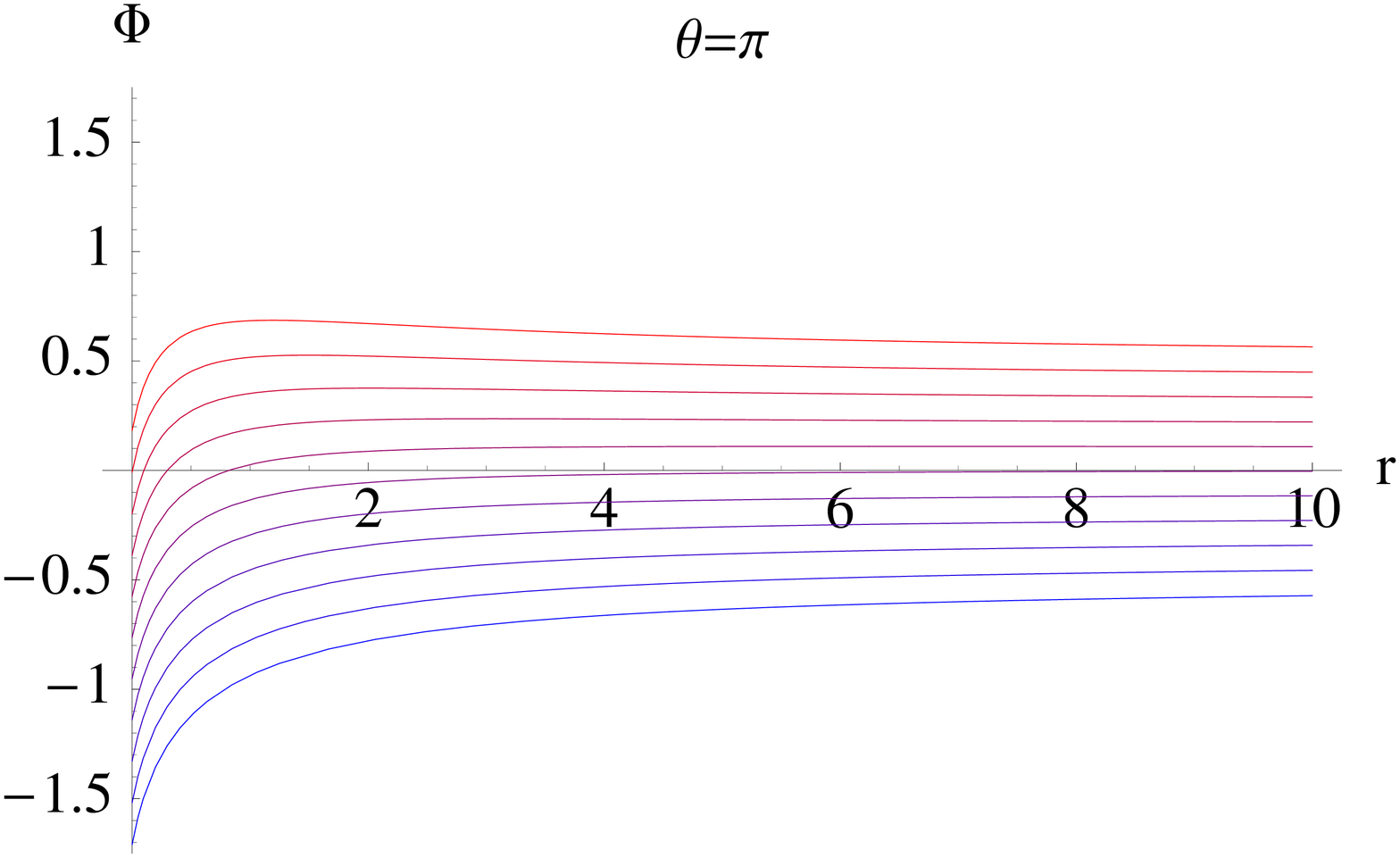}
\epsfysize=1.25in
\epsfbox{legend.eps}}
\end{center}
\caption{Radial evolution of the scalar field for an ergo-branch
black hole starting with different
asymptotic values at five different values of $\theta$.
We take
$P=Q= 2\sqrt\pi$ and $J=4\pi\sqrt{2}$
for $\Phi_\infty=0$, and then
change $\Phi_\infty$ and $P$, $Q$ using 
the transformation \refb{ekk2}.}
\label{f3}
\end{figure}

The scalar field $\Phi$ becomes in this limit
\be{escalarf}
e^{-4\Phi/\sqrt 3} = {f_p\over f_q}\, ,
\ee
 where $f_p$ and $f_q$ now refer to the functions $f_p$ and
 $f_q$ at the horizon:
  \be{esb6}
f_p = -M_K^2 \sin^2\theta + M_K \pp  +
\frac{\pp }{\pp +\q}\frac{\left(\pp -2M_{K}\right)\left(\q-2M_{K}\right)}{2}
 -\frac{\pp \sqrt{\left(\pp ^{2}-4M_{K}^{2}
\right)\left(\q^{2}-4M_{K}^{2}\right)}}{2(\pp +\q)}
\cos\theta \ee
\be{esb7}
f_q = -M_K^2 \sin^2\theta + M_K \q
+ \frac{\q}{\pp +\q}\frac{\left(
\pp -2M_{K}\right)\left(\q-2M_{K}\right)}{2}
+\frac{\q\sqrt{\left(\pp ^{2}-4M_{K}^{2}\right)\left(\q^{2}-4M_{K}^{2}
\right)}}{2 (\pp +\q)} \cos\theta  \, .
\ee
The near
horizon gauge field can also
be calculated by a tedious but straightforward
procedure after taking into account the change in coordinates described
above. The final result  is  of the form given in \refb{et2} with
\be{esb3a}
 e - \alpha b(\theta) =  {M_K \sqrt{\pp\q}\over 4\sqrt\pi\, f_q}
 \, \left( {1\over 2}\,
 {\pp\over \q}\, Q \, \sin^2\theta +P\, \sqrt{\q\over \pp} \,
 \cos\theta\right)\, . \ee
This gives
\be{eeexpnew}
e = {1\over 2} \left[(e-\alpha b(\pi)) + (e-\alpha b(0))\right]
= -{P^2 Q\over 4\sqrt\pi \sqrt{J^2 - P^2 Q^2}}\, ,
\ee
\be{epexpnew}
p = -{2\pi\over  \alpha} \left[(e-\alpha b(\pi)) - (e-\alpha b(0))\right]
= \sqrt\pi\, P\, ,
\ee
and
\be{qexpnew}
q = {8\pi\over \alpha} \, \left[ {e^{2\sqrt 3\Phi} b'\over
\sin\theta}\right]_0^\pi = 4\sqrt\pi\, Q\, .
\ee
Finally, the entropy associated with this solution can be easily calculated
by computing the area of the  horizon, and is given by
\be{esb8}
S_{BH} = 2\pi\,  \sqrt{J^2-P^2 Q^2 }\, .
\ee
We have explicitly checked that the near horizon
ergo-branch field configurations
described above satisfy the differential equations
\refb{dpsi}-\refb{eieqn}.

\begin{figure}
\leavevmode
\begin{center}
\hbox{\epsfysize=2in
\epsfbox{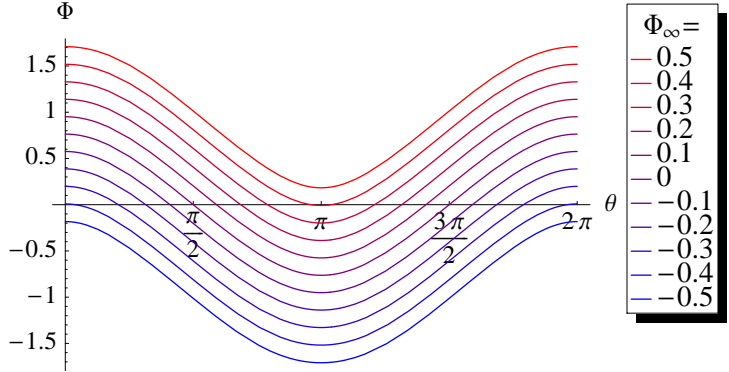}
\epsfysize=2in
\epsfbox{legend.eps}}
\end{center}
\caption{Scalar field profile at the horizon for a black hole on the
ergo-branch for different asymptotic values of $\Phi$.
We take
$P=Q= 2\sqrt\pi$ and $J=4\pi\sqrt{2}$
for $\Phi_\infty=0$, and then
change $\Phi_\infty$ and $P$, $Q$ using 
the transformation \refb{ekk2}.
Clearly the scalar field profile at
the horizon depends on its asymptotic
value.}
\label{f4}
\end{figure}

The entropy is clearly invariant under the transformation \refb{ekk2.5}.
However in this case the near horizon background is not invariant
under the transformation \refb{ekk2}. One way to see this is to note
that under the transformation \refb{ekk2.5} the combination
$M_K^2 \pp \q=(J^2-P^2 Q^2)/64\pi^2$ remains invariant.
This shows that $M_K$
cannot remain invariant under this transformation, since if $M_K$ had
been invariant then $\pp \q$ would be invariant, and the invariance of
$J$ given in \refb{esb2} would imply that $\pp +\q$ is also invariant.
This in turn would mean that $M_K$, $\pp $ and $\q$ are all
invariant under
\refb{ekk2.5} and hence $P$ and $Q$ would be invariant which is clearly
a contradiction. Given the fact that $M_K$ is not invariant under
this transformation we see that the coefficient of the $\sin^2\theta$ term
in $f_p$ and $f_q$ are not invariant under \refb{ekk2.5}. This in turn
shows that $\psi$, and hence
the background metric, is not invariant under the transformation
\refb{ekk2}.
This is also seen from figures \ref{f3} and \ref{f4}
where we have shown
respectively the radial evolution of the scalar field and the scalar field
profile at the horizon for different asymptotic values of $\Phi$.
Nevertheless several components of the near horizon
background, {\it e.g.} $\Omega(\theta)$ and the parameters $\alpha$
and $e$ do remain invariant under this transformation, indicating that
at least these components do get attracted towards fixed values as
we approach the horizon.

\subsection{Black Holes in Toroidally Compactified Heterotic
String Theory} \label{s3.2}

The theory under consideration
is a four dimensional theory of
gravity coupled to a complex scalar $S=S_1+iS_2$,
a 4$\times 4$
matrix valued scalar field $M$ satisfying the constraint
\be{es3.1}
MLM^T = L, \qquad L=\pmatrix{0 & I_2\cr I_2 & 0}\,,
\ee
and four U(1) gauge fields $A_\mu^{(i)}$
($1\le i\le 4$).\footnote{Actual heterotic string theory has 28 gauge
fields and a 28$\times$28 matrix valued scalar field, but the
truncated theory discussed here contains all the non-trivial information
about the theory.} Here $I_2$ denotes $2\times 2$ identity matrix.
The
bosonic part of the lagrangian density is
\bea{es3.2}
\LL &=& R - {1\over 2}
\, g^{\mu\nu} S_2^{-2} \p_\mu \bar S \p_\nu S  + {1\over 8} g^{\mu\nu}
Tr(\p_\mu M L \p_\nu M L) \nonumber \\
&& -
{1\over 4}\, S_2 g^{\mu\rho}
g^{\nu\sigma}
F^{(i)}_{\mu\nu} (LML)_{ij} F^{(j)}_{\rho\sigma}
+ {1\over 4}\, S_1 g^{\mu\rho}
g^{\nu\sigma}
F^{(i)}_{\mu\nu} L_{ij}
\wt F^{(j)}_{ \rho\sigma}\, ,
\eea
where
\be{defft}
\wt F^{(i)\mu\nu} = {1\over 2}\, (\sqrt{-\det g})^{-1}
\epsilon^{\mu\nu\rho\sigma} \, \wt F^{(i)}_{ \rho\sigma}\, .
\ee
General rotating black solution in this theory, carrying electric charge
vector $\vec q$ and magnetic charge vector $\vec p$, has been
constructed in \cite{9603147}. Before we begin analyzing the solution,
we would like to note that the lagrangian density \refb{es3.2}
is invariant under an SO(2,2) rotation:
\be{es3.3}
M\to \Omega M \Omega^T, \quad F^{(i)}_{\mu\nu}\to \Omega_{ij}
F^{(j)}_{\mu\nu}\,  ,
\ee
where $\Omega$ is a 4$\times$4 matrix satisfying
\be{es3.4}
\Omega L \Omega^T = L\, .
\ee
Thus given a classical solution, we can generate a class of classical
solutions using this transformation. Since the magnetic and electric
charges $p_i$ and $q_i$
are proportional to $F^{(i)}_{\theta\phi}$ and
$\p \LL / \p F^{(i)}_{rt}$ respectively, we see that under the
transformation
\refb{es3.3}, $p_i\to \Omega_{ij} p_j$,
$q_i\to (\Omega^T)^{-1}_{ij} q_j$.
Thus if we want the new solution to have the same electric and magnetic
charges, we must make compensating transformation in the parameters
labelling the electric and magnetic charges.
This shows that we can generate a family of solutions
carrying the same electric and magnetic charges by making the
transformation:
\be{es3.5}
M\to \Omega M \Omega^T, \quad F^{(i)}_{\mu\nu}\to \Omega_{ij}
F^{(j)}_{\mu\nu}, \quad Q_i \to \Omega^T_{ij} Q_j, \quad
P_i\to \Omega^{-1}_{ij} P_j\, ,
\ee
where $\vec Q$ and $\vec P$ are the parameters which label electric
and magnetic charges in the original solution. This transformation
changes the asymptotic value of $M$ leaving the charges
unchanged. Thus the general argument of section \ref{s0} will
imply that the entropy must remain invariant under
such a transformation. Invariance of the entropy under
the transformation \refb{es3.3}, which is a symmetry of the theory,
will then imply that the entropy must be invariant
under
\be{es3.5.5}
Q_i \to \Omega^T_{ij} Q_j, \quad
P_i\to \Omega^{-1}_{ij} P_j\, .
\ee
On the other hand if there is a unique background for a
given set of charges
then the background itself must be invariant under the transformation
\refb{es3.5}.

The equations of motion derived from the lagrangian density
\refb{es3.2} is also invariant under the electric magnetic duality
transformation:
\be{es3.6}
S\to {a S + b\over c S +d}\,  , \qquad F^{(i)}_{\mu\nu}\to
(c S_1 + d) F^{(i)}_{\mu\nu}  + c S_2 (ML)_{ij}
\wt F^{(j)}_{\mu\nu}\, ,
\ee
where $a,b,c,d$ are real numbers satisfying $ad-bc=1$.
We can use this transformation to generate a family of black hole
solutions from a given solution.
{}From the definition of electric and magnetic charges it follows
that under this transformation the electric and magnetic charge
vectors $\vec q$, $\vec p$ transform as:
\be{es3.7}
 \vec q\to (a\vec q
- b L \vec p),  \qquad \vec p\to (-c L\vec q + d\vec p)\, .
\ee
Thus if we want the new solution to have the same charges as the old
solution we must perform compensating transformation on the electric
and magnetic charge parameters $\vec Q$ and $\vec P$.
We can get a family of solutions with the
same electric and magnetic charges but different asymptotic values of
the scalar field $S$ by the transformation:
\be{es3.8}
S\to {a S + b\over c S +d}, \quad F^{(i)}_{\mu\nu}\to
(c S_1 + d) F^{(i)}_{\mu\nu}  + c S_2 (ML)_{ij}
\wt F^{(j)}_{\mu\nu}, \quad \vec Q\to d\vec Q + b L\vec P,
\quad \vec P\to cL\vec Q + a \vec P\, .
\ee
Arguments similar to the one given for the O(2,2) transformation
shows that the entropy
must remain invariant under the transformation
\be{es3.8.5}
\vec Q\to d\vec Q + b L\vec P,
\quad \vec P\to cL\vec Q + a \vec P\, .
\ee
Furthermore if the
entropy function has a unique extremum then the near horizon field
configuration must also remain invariant under the transformation
\refb{es3.8}.

\subsubsection{The black hole solution}

Ref.\cite{9603147} constructed
rotating black hole solutions in this theory
carrying the following charges:
\be{es3.9}
Q = \pmatrix{ 0 \cr Q_2 \cr 0 \cr Q_4}\, ,
\qquad P=\pmatrix{P_1 \cr 0 \cr P_3\cr
0}
\, .
\ee
These black holes break all the supersymmetries of the theory.
In order to describe the solution we parametrize the matrix valued scalar
field $M$ as
\be{es33.1}
M =\pmatrix{G^{-1} & -G^{-1} B\cr B G^{-1} & G - B G^{-1} B}
\ee
where $G$ and $B$ are $2\times 2$ matrices of the form
\be{es33.2}
G=\pmatrix{G_{11} & G_{12} \cr G_{12} & G_{22}}, \qquad
B=\pmatrix{0 & B_{12} \cr -B_{12} & 0}\, .
\ee
Physically $G$ and $B$ represent components of the string metric and
the anti-symmetric tensor field along an internal two dimensional
torus.
The solution is given by
\bea{es33.3}
G_{11}&=&{{(r+2m{\rm sinh}^2 \delta_{4})
(r+2m{\rm sinh}^2 \delta_{2})
+l^2{\rm cos}^2 \theta}\over {(r+2m{\rm sinh}^2 \delta_{3})(r+2m
{\rm sinh}^2 \delta_{2})+l^2{\rm cos}^2\theta}},  \cr
G_{12}&=&{2ml{\rm cos}
\theta({\rm sinh}\delta_{3}{\rm cosh}\delta_{4}
{\rm sinh}\delta_{1}{\rm cosh}\delta_{2}-{\rm cosh}\delta_{3}
{\rm sinh}\delta_{4}{\rm cosh}\delta_{1}{\rm sinh}\delta_{2})\over
{(r+2m{\rm sinh}^2 \delta_{3})(r+2m{\rm sinh}^2 \delta_{2})+
l^2{\rm cos}^2\theta}},  \cr
G_{22}&=&{{(r+2m{\rm sinh}^2 \delta_{3})(r+2m{\rm sinh}^2 \delta_{1})
+l^2{\rm cos}^2 \theta}\over {(r+2m{\rm sinh}^2 \delta_{3})(r+2m
{\rm sinh}^2 \delta_{2})+l^2{\rm cos}^2\theta}},  \cr
B_{12}&=&-{{2ml{\rm cos}\theta({\rm sinh}\delta_{3}{\rm cosh}
\delta_{4}{\rm cosh}\delta_{1}{\rm sinh}\delta_{2}-{\rm cosh}
\delta_{3}{\rm sinh}\delta_{4}{\rm sinh}\delta_{1}{\rm cosh}
\delta_{2})}\over{(r+2m{\rm sinh}^2 \delta_{3})(r+2m{\rm sinh}^2
\delta_{2})+l^2{\rm cos}^2\theta}},  \cr
Im\, S&=&{\Delta^{1\over 2}\over
{(r+2m{\rm sinh}^2 \delta_{3})(r+2m{\rm sinh}^2
\delta_{4})+l^2{\rm cos}^2 \theta}},\cr
ds^2&=&\Delta^{1\over 2}[-{{r^2-2mr+l^2{\rm cos}^2\theta}\over
\Delta}dt^2+{{dr^2}\over{r^2-2mr+l^2}} + d\theta^2
+{{{\rm sin}^2\theta}\over \Delta}\{(r+2m{\rm sinh}^2
\delta_{3})\cr
&\times&(r+2m{\rm sinh}^2 \delta_{4})(r+2m{\rm sinh}^2 \delta_{1})
(r+2m{\rm sinh}^2 \delta_{2})+l^2(1+{\rm cos}^2\theta)r^2+W \cr
&+&2ml^2r{\rm sin}^2\theta\}d\phi^2
-{{4ml}\over \Delta}\{({\rm cosh} \delta_{3}{\rm cosh}
\delta_{4}{\rm cosh} \delta_{1}{\rm cosh} \delta_{2} \cr
&-&{\rm sinh} \delta_{3}{\rm sinh} \delta_{4}
{\rm sinh} \delta_{1}{\rm sinh} \delta_{2})r
+2m{\rm sinh}\delta_{3}{\rm sinh}\delta_{4}{\rm sinh}\delta_{1}
{\rm sinh}\delta_{2}\}{\rm sin}^2 \theta dtd\phi], \nonumber \\
 \eea
where
\bea{es33.4}
\Delta &\equiv& (r+2m{\rm sinh}^2 \delta_{3})
(r+2m{\rm sinh}^2 \delta_{4})(r+2m{\rm sinh}^2 \delta_{1})
(r+2m{\rm sinh}^2 \delta_{2})\cr
&+&(2l^2r^2+W){\rm cos}^2\theta, \cr
W &\equiv& 2ml^2({\rm sinh}^2\delta_{3}+{\rm sinh}^2\delta_{4}+
{\rm sinh}^2\delta_{1}+{\rm sinh}^2\delta_{2})r \cr
&+&4m^2l^2(2{\rm cosh}\delta_{3}{\rm cosh}\delta_{4}{\rm cosh}
\delta_{1}{\rm cosh}\delta_{2}{\rm sinh}\delta_{3}{\rm sinh}
\delta_{4}{\rm sinh}\delta_{1}{\rm sinh}\delta_{2}\cr
&-&2{\rm sinh}^2 \delta_{3}{\rm sinh}^2 \delta_{4}{\rm sinh}^2
\delta_{1}{\rm sinh}^2 \delta_{2}-{\rm sinh}^2 \delta_{4}
{\rm sinh}^2 \delta_{1}{\rm sinh}^2 \delta_{2} \cr
&-&{\rm sinh}^2 \delta_{3}{\rm sinh}^2 \delta_{1}
{\rm sinh}^2 \delta_{2}-{\rm sinh}^2 \delta_{3}{\rm sinh}^2
\delta_{4}{\rm sinh}^2 \delta_{2}-{\rm sinh}^2 \delta_{3}
{\rm sinh}^2 \delta_{4}{\rm sinh}^2 \delta_{1})\cr
&+&l^4{\rm cos}^2 \theta.
\eea
$a$, $m$, $\delta_{1}$, $\delta_{2}$, $\delta_{3}$ and
$\delta_{4}$ are parameters labelling the solution.
Ref.\cite{9603147} did not explicitly present the results for $Re~S$
and the gauge fields.

The ADM mass $M$,  electric and magnetic charges $\{Q_i,P_i\}$,
and the angular momentum $J$ are given by:\footnote{In
defining $M$ and $J$ we
have taken into
account our convention
$G_N=16\pi$, and also the fact that our definition of the angular
momentum differs from the standard one by a minus sign.
Normalizations
of $\vec Q$ and $\vec P$ are arbitrary at this stage.}
\bea{es33.5}
M&=&8\pi\, m({\rm cosh}^2 \delta_{1}+{\rm cosh}^2 \delta_{2}+
{\rm cosh}^2 \delta_{3}+{\rm cosh}^2 \delta_{4})-16\pi\, m, \cr
Q_2 &=&4\sqrt{2\pi}\, m\,
{\rm cosh}\delta_{1}{\rm sinh}\delta_{1},\ \ \ \
\
Q_4 = 4\sqrt{2\pi} \, m\,
{\rm cosh}\delta_{2}{\rm sinh}\delta_{2}, \cr
P_1 &=& 4\sqrt{2\pi}
\, m\, {\rm cosh}\delta_{3}{\rm sinh}\delta_{3},\ \ \ \ \
P_3 = 4\sqrt{2\pi}
\, m\, {\rm cosh}\delta_{4}{\rm sinh}\delta_{4}, \cr
J&=&-16\pi\,
lm({\rm cosh}\delta_{1}{\rm cosh}\delta_{2}{\rm cosh}\delta_{3}
{\rm cosh}\delta_{4}-{\rm sinh}\delta_{1}{\rm sinh}\delta_{2}
{\rm sinh}\delta_{3}{\rm sinh}\delta_{4}).
\eea
The entropy associated with this solution was computed in
\cite{9603147} to be
\be{es33.6}
S_{BH}=32\pi^2 \Big[m^2(\prod^4_{i=1}\cosh\delta_i+\prod^4_{i=1}
\sinh\delta_i)+m\sqrt{m^2-l^2}(\prod^4_{i=1}
\cosh\delta_i-\prod^4_{i=1}\sinh\delta_i)\Big] \, .
\ee

As in the case of Kaluza-Klein black hole this solution also has two
different kinds of extremal limit which we shall denote by ergo-branch
and ergo-free branch. The ergo-branch was discussed in \cite{9603147}.

 \subsubsection{The ergo-branch}

 The extremal limit corresponding to the ergo-branch is obtained by
 taking the limit $l\to m$.
In this limit the second term in the
 expression for the entropy vanishes and the first term gives
\be{es3.9.5}
S_{BH} = 2\pi \, \sqrt{J^2 + Q_2 Q_4 P_1 P_3}\, .
\ee
Now the most general  transformation of the form \refb{es3.5.5}
which does not take the
charges given in \refb{es3.9} outside this family is:
\be{es3.10}
\Omega = \pmatrix{e^\gamma & 0 & 0 & 0\cr 0 & e^\beta & 0 & 0\cr
0 & 0 & e^{-\gamma} & 0\cr 0 & 0 & 0 & e^{-\beta}}\, ,
\ee
for real parameters $\gamma$, $\beta$. This gives
\be{newpqtr}
P_1\to e^{-\gamma} P_1, \quad P_3\to e^{\gamma} P_3,
\quad Q_2\to e^\beta Q_2, \quad Q_4\to e^{-\beta} Q_4\, .
\ee
On the other hand most general transformation of the type \refb{es3.8.5}
which keeps the charge vector within the same family is
\be{es3.11}
\pmatrix{a & b\cr c & d} = \pmatrix{a & 0\cr 0 & a^{-1}}\, .
\ee
This gives
\be{newpqtrs}
P_1\to   a\, P_1, \quad P_3\to a\,  P_3,
\quad Q_2\to a^{-1}\,  Q_2, \quad Q_4\to a^{-1}\,  Q_4\, .
\ee
It is easy to see that the entropy
given in \refb{es3.9.5} does not change under the transformations
\refb{newpqtr}, \refb{newpqtrs}.\footnote{As in
\refb{epexp},
\refb{qexp}, the parameters
$\vec P$,
$\vec Q$ are related to the charges $\vec p$, $\vec q$
by some overall
normalization factors. These factors
do not affect the transformation laws
of the charges given in \refb{newpqtr}, \refb{newpqtrs}.}

After some tedious manipulations along the lines described in
section \ref{sbranch}, the near horizon metric can be brought into
the form given in eq.\refb{et2} with
\bea{eneh}
&& \Omega = {1\over 8\pi} \sqrt{J^2 + Q_2 Q_4 P_1 P_3}\, \sin\theta,
\qquad e^{-2\psi} = {1\over 64\pi^2} \, (J^2 + Q_2 Q_4 P_1 P_3)\,
\sin^2\theta\, \Delta^{-1/2}\, , \nonumber \\
&& \qquad \alpha = {J\over
\sqrt{J^2 + Q_2
Q_4 P_1 P_3}}\, ,
\eea
where $\Delta$ has to be evaluated on the horizon $r=m$.
We have found that the near horizon
metric and the scalar fields are
not invariant under the corresponding transformations
\refb{es3.5} and \refb{es3.8} generated by the matrices
\refb{es3.10} and \refb{es3.11} respectively,
essentially due to the fact that
$\Delta$ is not invariant under these transformations. This shows
that in this case for a fixed set of
charges the entropy function has a family of extrema.

\subsubsection{The ergo-free branch}

The extremal limit in the ergo-free branch is obtained by taking one
or three
of the $\delta_i$'s negative, and then taking the limit $|\delta_i|\to
\infty$, $m\to 0$, $l\to 0$ in a way that keeps the $Q_i$, $P_i$ and
$J$ finite. It is easy to see that in this limit the first term in the
expression \refb{es33.6}
for the entropy vanishes and the second term
gives\footnote{Note that the product $Q_2Q_4 P_1 P_3$
is negative due to the fact that
an odd number of $\delta_i$'s are negative.}
\be{es3.9.6}
S_{BH} = 2\pi \, \sqrt{-J^2 - Q_2 Q_4 P_1 P_3}\, .
\ee
Again we see that $S_{BH}$ is invariant under the transformations
\refb{newpqtr}, \refb{newpqtrs}.

On the ergo-free branch the horizon is at $r=0$.
The near horizon background
can be computed easily from \refb{es33.3} following the approach
described in section \ref{snearh}
and has the following form after appropriate rescaling of the
time coordinate:
\bea{es34.1}
ds^2 &=&  {1\over 8\pi} \sqrt{-Q_2Q_4P_1P_3-J^2\cos^2\theta}
\left(-r^{2}dt^{2}
+\frac{dr^{2}}{r^{2}}+d\theta^{2}\right) \nonumber \\
&&
+ {1\over 8\pi} \, {-Q_2Q_4P_1P_3-J^2\over
\sqrt{-Q_2Q_4P_1P_3-J^2\cos^2\theta}}\,
\sin^2\theta\, (d\phi -\alpha r dt)^2\, ,
\eea
\be{es34.1.5}
Im S = \sqrt{-{Q_2 Q_4\over P_1 P_3} - {J^2 \cos^2\theta
\over (P_1P_3)^2}}\, ,
\ee
\be{es34.2}
G_{11} = \left|{P_3\over P_1}\right|, \qquad G_{12}=
-{J\cos\theta\over P_1
Q_2}\left|{Q_2
\over Q_4}\right|, \qquad
G_{22}= \left|{Q_2\over Q_4}\right|, \qquad B_{12} =
{J\cos\theta\over P_1 Q_4}
\, ,
\ee
where
\be{es34.3}
\alpha = -J / \sqrt{-Q_2Q_4P_1P_3-J^2}\, .
\ee
It is easy to see that the background is invariant under
\refb{es3.5} and \refb{es3.8} for transformation matrices
of the form
described in \refb{es3.10} and \refb{es3.11}.

\subsubsection{Duality invariant form of the entropy}

In the theory described here a combination of the charges that is
invariant under both transformations \refb{es3.5.5} and \refb{es3.8.5}
is
\be{edefd}
D \equiv
(Q_1 Q_3 + Q_2 Q_4) (P_1 P_3 + P_2 P_4) -
{1\over 4} (Q_1 P_1 + Q_2 P_2 + Q_3 P_3 + Q_4 P_4)^2 \, .
\ee
Thus we expect the entropy to depend on the charges through this
combination. Now for the charge vectors given in \refb{es3.9} we
have
\be{edu1}
D = Q_2 Q_4 P_1 P_3\, .
\ee
Using this result
we can express the entropy formula \refb{es3.9.5} in the ergo-branch
in the duality invariant form\cite{9603147}:
\be{edu2}
S_{BH} = 2\pi\, \sqrt{J^2 + D}\, .
\ee
On the other hand the formula \refb{es3.9.6}
on the ergo-free branch may be
expressed as
\be{edu3}
S_{BH} = 2\pi\, \sqrt{-J^2 - D}\, .
\ee

We now note that the Kaluza-Klein black hole described in section
\refb{s3.1} also falls into the general class of black holes discussed
in this section with charges:
\be{edu4}
Q = \sqrt{2}\, \pmatrix{Q\cr 0\cr 0\cr 0}, \qquad
P=\sqrt{2}\, \pmatrix{P \cr 0\cr 0\cr 0}\, .
\ee
Thus in this case
\be{edu5}
D = -P^2 Q^2\, .
\ee
We can now recognize the entropy formul\ae\ \refb{entr1} and
\refb{esb8} as special cases of \refb{edu3} and \refb{edu2}
respectively.

Finally we can try to write down the
near horizon metric on the ergo-free branch
in a form that holds for the black hole
solutions analyzed in this as well as in
the previous subsection and which
makes
manifest the
invariance of the background under arbitrary transformations of the form
described in \refb{es3.5}, \refb{es3.8}. This is of the form:
\bea{edu6}
ds^2 &=& {1\over 8\pi} \sqrt{-D-J^2\cos^2\theta}
\left(-r^{2}dt^{2}
+\frac{dr^{2}}{r^{2}}+d\theta^{2}\right) \nonumber \\
&&
+ {1\over 8\pi} \, {-D-J^2\over
\sqrt{-D-J^2\cos^2\theta}} \, \sin^2\theta\,
(d\phi -\alpha r dt)^2\, ,
\eea
where
\be{edu8}
\alpha = -{J\over \sqrt{-D-J^2}}\, .
\ee
\refb{edd1} and \refb{es34.1} are special cases of this equation.

\bigskip

{\bf Acknowledgment:} We would like to thank A.~Dabholkar and
S.~Minwalla for useful discussions. This work was supported by
Department of Atomic Energy, Government of India. SPT acknowledges
further support from the Swarnajayanti Fellowship, DST, Government
of India. DA and KG would like to thank ICTP, Trieste for
hospitality. We thank the people of India for generously supporting
research in String Theory.


\begin{thebibliography}{99}

\bibitem{9508072}
S.~Ferrara, R.~Kallosh and A.~Strominger,
``N=2 extremal black holes,''
Phys.\ Rev.\ D {\bf 52}, 5412 (1995)
[arXiv:hep-th/9508072].

\bibitem{9602111}
A.~Strominger,
``Macroscopic Entropy of $N=2$ Extremal Black Holes,''
Phys.\ Lett.\ B {\bf 383}, 39 (1996)
[arXiv:hep-th/9602111].

\bibitem{9602136}
S.~Ferrara and R.~Kallosh,
``Supersymmetry and Attractors,''
Phys.\ Rev.\ D {\bf 54}, 1514 (1996)
[arXiv:hep-th/9602136].

\bibitem{0507096}
  K.~Goldstein, N.~Iizuka, R.~P.~Jena and S.~P.~Trivedi,
  ``Non-supersymmetric attractors,''
  arXiv:hep-th/0507096.

  \bibitem{0510024}
  R.~Kallosh,
  ``New attractors,''
  arXiv:hep-th/0510024.

 \bibitem{0511117}
  P.~K.~Tripathy and S.~P.~Trivedi,
  ``Non-supersymmetric attractors in string theory,''
  arXiv:hep-th/0511117.

\bibitem{0511215}
  A.~Giryavets,
  ``New attractors and area codes,''
  arXiv:hep-th/0511215.

\bibitem{0512138}
  K.~Goldstein, R.~P.~Jena, G.~Mandal and S.~P.~Trivedi,
  ``A C-Function For Non-Supersymmetric Attractors,''
  arXiv:hep-th/0512138.

  \bibitem{0602005}
  R.~Kallosh, N.~Sivanandam and M.~Soroush,
  ``The non-BPS black hole attractor equation,''
  arXiv:hep-th/0602005.

\bibitem{0603003}
  R.~Kallosh,
  ``From BPS to Non-BPS Black Holes Canonically,''
  arXiv:hep-th/0603003.

  \bibitem{0511306}
  P.~Prester,
  ``Lovelock type gravity and
  small black holes in heterotic string theory,''
  JHEP {\bf 0602}, 039 (2006)
  [arXiv:hep-th/0511306].

  \bibitem{0601016}
  M.~Alishahiha and H.~Ebrahim,
  ``Non-supersymmetric attractors and entropy function,''
  arXiv:hep-th/0601016.

\bibitem{0601183}
  A.~Sinha and N.~V.~Suryanarayana,
  ``Extremal single-charge small black holes: Entropy function analysis,''
  arXiv:hep-th/0601183.

 \bibitem{0602022}
  B.~Chandrasekhar, S.~Parvizi, A.~Tavanfar and H.~Yavartanoo,
  ``Non-supersymmetric attractors in R**2 gravities,''
  arXiv:hep-th/0602022.

\bibitem{9711053}
J.~M.~Maldacena, A.~Strominger and E.~Witten,
``Black hole entropy in M-theory,''
JHEP {\bf 9712}, 002 (1997)
[arXiv:hep-th/9711053].


\bibitem{9801081}
K.~Behrndt, G.~Lopes Cardoso, B.~de Wit,
D.~Lust, T.~Mohaupt and W.~A.~Sabra,
 ``Higher-order black-hole solutions in N = 2 supergravity and Calabi-Yau
string backgrounds,''
Phys.\ Lett.\ B {\bf 429}, 289 (1998) [arXiv:hep-th/9801081].

\bibitem{9812082}
G.~Lopes Cardoso, B.~de Wit and T.~Mohaupt,
``Corrections to macroscopic supersymmetric black-hole entropy,''
Phys.\ Lett.\ B {\bf 451}, 309 (1999)
[arXiv:hep-th/9812082].

\bibitem{9904005}
G.~Lopes Cardoso, B.~de Wit and T.~Mohaupt,
``Deviations from the area law for supersymmetric black holes,''
Fortsch.\ Phys.\  {\bf 48}, 49 (2000)
[arXiv:hep-th/9904005].

\bibitem{9906094}
G.~Lopes Cardoso, B.~de Wit and T.~Mohaupt,
``Macroscopic entropy formulae and non-holomorphic corrections for
supersymmetric black holes,''
Nucl.\ Phys.\ B {\bf 567}, 87 (2000)
[arXiv:hep-th/9906094].

\bibitem{9910179}
G.~Lopes Cardoso, B.~de Wit and T.~Mohaupt,
``Area law corrections from state counting and supergravity,''
Class.\ Quant.\ Grav.\  {\bf 17}, 1007 (2000)
[arXiv:hep-th/9910179].

\bibitem{0007195}
T.~Mohaupt,
``Black hole entropy, special geometry and strings,''
Fortsch.\ Phys.\  {\bf 49}, 3 (2001)
[arXiv:hep-th/0007195].

\bibitem{0009234}
G.~Lopes Cardoso, B.~de Wit, J.~Kappeli and T.~Mohaupt,
``Stationary BPS solutions in N = 2 supergravity with R**2
interactions,''
JHEP {\bf 0012}, 019 (2000)
[arXiv:hep-th/0009234].

\bibitem{0012232}
G.~L.~Cardoso, B.~de Wit, J.~Kappeli and T.~Mohaupt,
``Examples of stationary BPS solutions in N = 2 supergravity theories
with
R**2-interactions,''
Fortsch.\ Phys.\  {\bf 49}, 557 (2001)
[arXiv:hep-th/0012232].

\bibitem{0409148}
A.~Dabholkar,
``Exact counting of black hole microstates,''
arXiv:hep-th/0409148.

\bibitem{0410076}
A.~Dabholkar, R.~Kallosh and A.~Maloney,
``A stringy cloak for a classical singularity,''
arXiv:hep-th/0410076.

\bibitem{0411255}
A.~Sen,
``How does a fundamental string stretch its horizon?,''
arXiv:hep-th/0411255.

\bibitem{0411272}
V.~Hubeny, A.~Maloney and M.~Rangamani,
``String-corrected black holes,''
arXiv:hep-th/0411272.

\bibitem{0501014}
D.~Bak, S.~Kim and S.~J.~Rey,
``Exactly soluble BPS black holes in higher curvature N = 2
supergravity,''
arXiv:hep-th/0501014.

\bibitem{0506176}
  P.~Kraus and F.~Larsen,
  ``Microscopic black hole entropy in theories with higher derivatives,''
  arXiv:hep-th/0506176.

  \bibitem{0506177}
  A.~Sen,
  ``Black hole entropy function and the attractor mechanism in higher
  derivative gravity,''
  arXiv:hep-th/0506177.

\bibitem{0508042}
A.~Sen,
``Entropy function for heterotic black holes,''
arXiv:hep-th/0508042.

\bibitem{0508218}
  P.~Kraus and F.~Larsen,
    arXiv:hep-th/0508218.

  \bibitem{0603149}
  B.~Sahoo and A.~Sen,
  ``Higher derivative corrections to
  non-supersymmetric extremal black holes in
  N = 2 supergravity,''
  arXiv:hep-th/0603149.

  \bibitem{0503219}
  P.~Kraus and F.~Larsen,
  ``Attractors and black rings,''
  Phys.\ Rev.\ D {\bf 72}, 024010 (2005)
  [arXiv:hep-th/0503219].

  \bibitem{0601228}
  B.~Sahoo and A.~Sen,
  ``BTZ black hole with Chern-Simons and higher derivative terms,''
  arXiv:hep-th/0601228.

\bibitem{0602292}
  S.~Parvizi and A.~Tavanfar,
  ``Partition function of non-supersymmetric
black holes in the
supergravity
  limit,''
  arXiv:hep-th/0602292.

\bibitem{0605279}
  M.~Alishahiha and H.~Ebrahim,
  ``New attractor, entropy function
and black hole partition function,''
  arXiv:hep-th/0605279.

\bibitem{0606108}
S.~Ferrara and M.~Gunaydin,
``Orbits and Attractors for N=2 Maxwell-Einstein
Supergravity Theories in Five Dimensions''
  arXiv:hep-th/0606108.


\bibitem{0604106}
  A.~Ghodsi,
   ``R**4 corrections to D1D5p black hole entropy from entropy function
  arXiv:hep-th/0604106.



\bibitem{9602065}
  J.~C.~Breckenridge, R.~C.~Myers, A.~W.~Peet and C.~Vafa,
  ``D-branes and spinning black holes,''
  Phys.\ Lett.\ B {\bf 391}, 93 (1997)
  [arXiv:hep-th/9602065].

\bibitem{9611094}
  R.~Kallosh, A.~Rajaraman and W.~K.~Wong,
  ``Supersymmetric rotating black holes and attractors,''
  Phys.\ Rev.\ D {\bf 55}, 3246 (1997)
  [arXiv:hep-th/9611094].

\bibitem{0605139}
  W.~Li and A.~Strominger,
  ``Supersymmetric probes in a rotating 5D attractor,''
  arXiv:hep-th/0605139.

\bibitem{9905099}
  J.~M.~Bardeen and G.~T.~Horowitz,
  ``The extreme Kerr throat geometry: A vacuum analog of AdS(2) x S(2),''
  Phys.\ Rev.\ D {\bf 60}, 104030 (1999)
  [arXiv:hep-th/9905099].

\bibitem{9307038}
R.~M.~Wald,
``Black hole entropy in the Noether charge,''
Phys.\ Rev.\ D {\bf 48}, 3427 (1993)
[arXiv:gr-qc/9307038].

\bibitem{9312023}
 T.~Jacobson, G.~Kang and R.~C.~Myers,
 ``On black hole entropy,''
Phys.\ Rev.\ D {\bf 49}, 6587 (1994) [arXiv:gr-qc/9312023].


\bibitem{9403028}
V.~Iyer and R.~M.~Wald,
``Some properties of Noether charge and a proposal for dynamical black
hole
entropy,''
Phys.\ Rev.\ D {\bf 50}, 846 (1994)
[arXiv:gr-qc/9403028].

\bibitem{9502009}
T.~Jacobson, G.~Kang and R.~C.~Myers,
``Black hole entropy in higher curvature gravity,''
arXiv:gr-qc/9502009.

 \bibitem{9505038}
D.~Rasheed, ``The rotating dyonic black holes of kaluza-klein theory,'' {\em
  Nucl. Phys.} {\bf B454} (1995) 379--401,
[arXiv:hep-th/9505038].

\bibitem{9610013}
T.~Matos and C.~Mora,
``Stationary dilatons with arbitrary electromagnetic
  field,'' {\em Class. Quant. Grav.} {\bf 14} (1997) 2331--2340,
[arXiv:hep-th/9610013].



\bibitem{9909102}
F.~Larsen, ``Rotating kaluza-klein black holes,''
{\em Nucl. Phys.} {\bf B575}
  (2000) 211--230,
[arXiv:hep-th/9909102].

\bibitem{9603147}
  M.~Cvetic and D.~Youm,
 ``Entropy of Non-Extreme
Charged Rotating Black Holes in String Theory,''
  Phys.\ Rev.\ D {\bf 54}, 2612 (1996)
  [arXiv:hep-th/9603147].

\bibitem{9601118}
  D.~P.~Jatkar, S.~Mukherji and S.~Panda,
  ``Rotating Dyonic Black Holes in Heterotic String Theory,''
  Phys.\ Lett.\ B {\bf 384}, 63 (1996)
  [arXiv:hep-th/9601118].

\end{thebibliography}
\end{document}